\documentclass[manuscript]{acmart}

\copyrightyear{2026}
\acmYear{2026}
\setcopyright{rightsretained}
\acmConference[EICS '26]{ACM Symposium on Engineering Interactive Computing Systems}{April 13--April 17, 2026}{Patras, Greece}
\acmBooktitle{ACM Symposium on Engineering Interactive Computing Systems (EICS '26), June 30--July 3, 2026, Patras, Greece}

\graphicspath{{figures/}{pictures/}{images/}{./}}

\usepackage{acro}
\DeclareAcronym{llm}{
    short=LLM,
    long=Large Language Model,
}
\DeclareAcronym{mvc}{
    short=MVC,
    long=Model-View-Controller,
}
\DeclareAcronym{bdar}{
    short=BDAR,
    long=Bidirectional Domain-Adaptive Representation,
}
\DeclareAcronym{cmda}{
    short=CMDA,
    long=Context-mediated domain adaptation,
}
\DeclareAcronym{cvc}{
    short=CVC,
    long=Context-View-Controller,
}
\DeclareAcronym{aco}{
    short=ACO,
    long=Adaptive Context Object,
}
\DeclareAcronym{ai}{
    short=AI,
    long=Artificial Intelligence,
}
\DeclareAcronym{ml}{
    short=ML,
    long=Machine Learning,
}
\DeclareAcronym{hci}{
    short=HCI,
    long=Human-Computer Interaction,
}
\DeclareAcronym{api}{
    short=API,
    long=Application Programming Interface,
}
\DeclareAcronym{iui}{
    short=IUI,
    long=Intelligent User Interfaces,
}
\DeclareAcronym{acm}{
    short=ACM,
    long=Association for Computing Machinery,
}

\usepackage{caption}
\usepackage{subcaption}
\usepackage{makecell} 

\usepackage[fixed]{fontawesome5}
\usepackage{xcolor}
\usepackage{xspace}
\usepackage{framed} 
\usepackage{changepage} 

\usepackage{listings}

\lstdefinelanguage{JavaScript}{
  morekeywords={break,case,catch,continue,debugger,default,delete,do,else,
    finally,for,function,if,in,instanceof,new,return,switch,this,throw,try,typeof,var,void,while,with,
    let,const,await,async,yield,of,import,export,from,class,extends,super,static,get,set},
  sensitive=true,
  morecomment=[l]{//},
  morecomment=[s]{/*}{*/},
  morestring=[b]',
  morestring=[b]",
  morestring=[b]`,
}

\definecolor{directedit}{HTML}{1447E6}
\definecolor{promptgeneration}{HTML}{8200DB}
\definecolor{contextgeneration}{HTML}{BB4D00}

\newcommand{\interactionmodechip}[2]{%
  {\setlength{\fboxsep}{2pt}%
    \colorbox{#1!5!white}{\sffamily\strut #2}}%
}

\newcommand{\directmanipulation}{%
  \interactionmodechip{directedit}{\textcolor{directedit}{\faIcon{pencil-alt}}~Direct Manipulation}\xspace}
  
\newcommand{\directedittext}[1]{%
  \interactionmodechip{directedit}{\textcolor{directedit}{#1}}}

\newcommand{\promptbasedregeneration}{%
  \interactionmodechip{promptgeneration}{\textcolor{promptgeneration}{\faIcon{comment}}~Prompt-based Regeneration}\xspace}

\newcommand{\promptedittext}[1]{%
  \interactionmodechip{promptgeneration}{\textcolor{promptgeneration}{#1}}}

\newcommand{\contextbasedgeneration}{%
  \interactionmodechip{contextgeneration}{\textcolor{contextgeneration}{\faIcon{star}}~Context-based Generation}\xspace}

\newcommand{\contextedittext}[1]{%
\interactionmodechip{contextgeneration}{\textcolor{contextgeneration}{#1}}}

\definecolor{expertcolor}{RGB}{0, 0, 128}  

\newenvironment{hquote}[1]{%
  \def\FrameCommand{%
    \hspace{1pt}%
    {\color{#1!35}\vrule width 2pt}%
    {\color{#1!35}\vrule width 4pt}%
    \colorbox{#1!10}%
  }%
  \MakeFramed{\advance\hsize-\width\FrameRestore}%
  \noindent\hspace{-4.55pt}
  \begin{adjustwidth}{}{7pt}%
  \vspace{2pt}\vspace{2pt}%
}
{%
  \vspace{2pt}\end{adjustwidth}\endMakeFramed%
}

\begin{document}


\title{Context-Mediated Domain Adaptation in Multi-Agent Sensemaking Systems}

\author{Anton Wolter}
\email{wol@cs.au.dk}
\orcid{0009-0004-6312-3355}
\affiliation{
  \institution{Aarhus University}
  \city{Aarhus}
  \country{Denmark}
}

\author{Leon Haag}
 \email{l.haag@alumni.maastrichtuniversity.nl}
 \orcid{0009-0005-4953-098X}
 \affiliation{
   \institution{Maastricht University}
   \city{Maastricht}
   \country{Netherlands}
}

\author{Vaishali Dhanoa}
\email{dhanoa@cs.au.dk}
\orcid{0000-0002-0493-8616}
\affiliation{
  \institution{Aarhus University}
  \city{Aarhus}
  \country{Denmark}
}
\affiliation{
  \institution{TU Wien}
  \city{Vienna}
  \country{Austria}
}

\author{Niklas Elmqvist}
\email{elm@cs.au.dk}
\orcid{0000-0001-5805-5301}
\affiliation{
  \institution{Aarhus University}
  \city{Aarhus}
  \country{Denmark}
}

\renewcommand{\shortauthors}{Wolter et al.}

\begin{abstract}
    Domain experts possess tacit knowledge that they cannot easily articulate through explicit specifications.
When experts modify AI-generated artifacts by correcting terminology, restructuring arguments, and adjusting emphasis, these edits reveal domain understanding that remains latent in traditional prompt-based interactions.
Current systems treat such modifications as endpoint corrections rather than as implicit specifications that could reshape subsequent reasoning.
We propose \textit{context-mediated domain adaptation}, a paradigm where user modifications to system-generated artifacts serve as implicit domain specification that reshapes \ac{llm}-powered multi-agent reasoning behavior.
Through our system \textsc{Seedentia}, a web-based multi-agent framework for sense-making, we demonstrate bidirectional semantic links between generated artifacts and system reasoning.
Our approach enables \textit{specification bootstrapping} where vague initial prompts evolve into precise domain specifications through iterative human-AI collaboration, implicit knowledge transfer through reverse-engineered user edits, and in-context learning where agent behavior adapts based on observed correction patterns.
We present results from an evaluation with domain experts who generated and modified research questions from academic papers.
Our system extracted 46 domain knowledge entries from user modifications, demonstrating the feasibility of capturing implicit expertise through edit patterns, though the limited sample size constrains conclusions about systematic quality improvements.

\end{abstract}


\setcopyright{cc}
\setcctype{by}
\acmJournal{PACMHCI}
\acmYear{2026} \acmVolume{10} \acmNumber{4} \acmArticle{EICS003}
\acmMonth{6} \acmDOI{10.1145/3812772}
 


\begin{teaserfigure}
  \includegraphics[width=\textwidth]{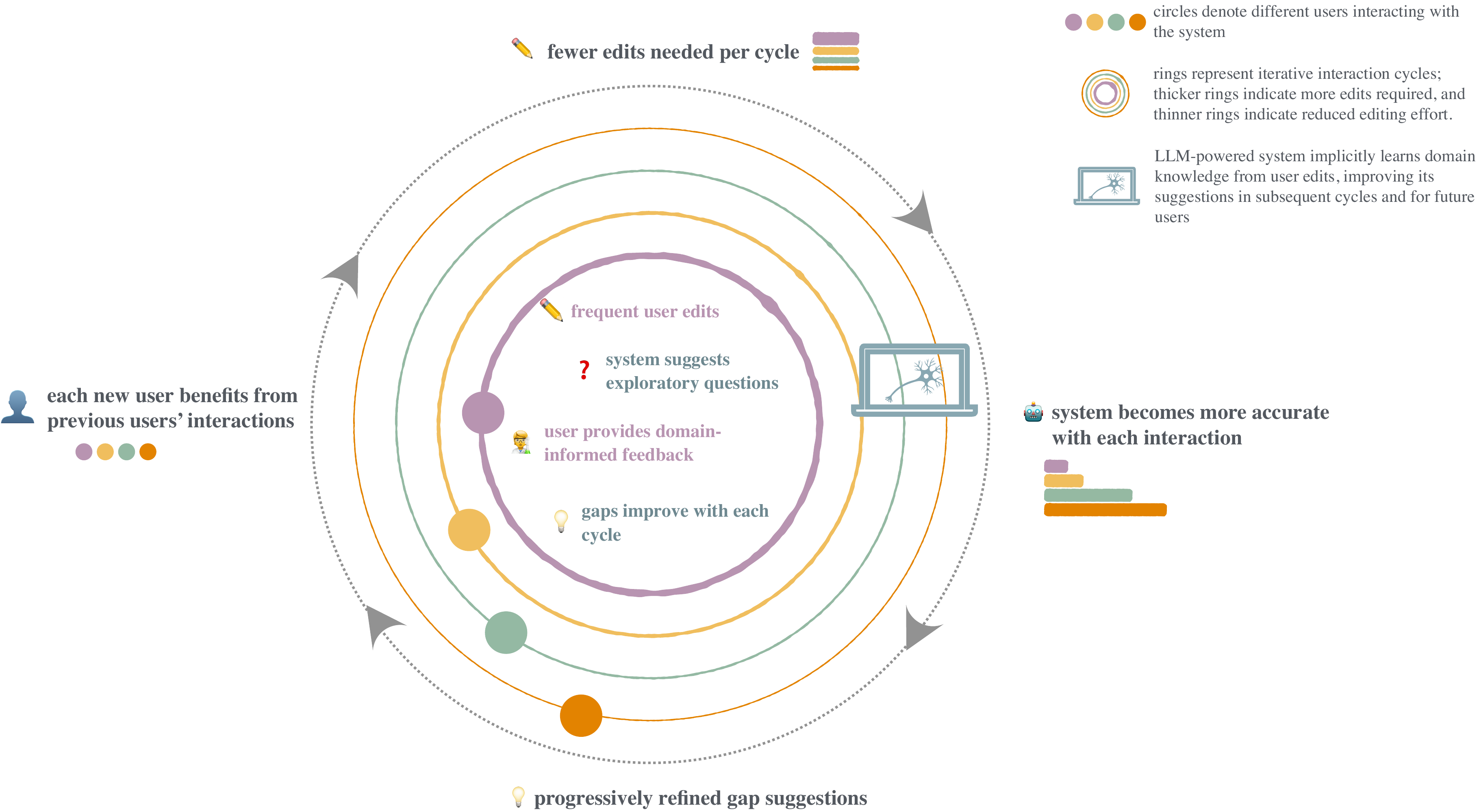}
  \caption{\textbf{Context-Mediated Domain Adaptation transforms ephemeral user interactions into persistent domain knowledge.}
  As user knowledge and LLM model knowledge deviate we analyze user interaction and edits in order to extract implicit domain knowledge.
  Through iterative refinement our approach expands the shared context substantially, capturing domain-specific terminology, conventions, and patterns.
  This accumulated knowledge persists in an LLM-agnostic format, enabling system improvements across sessions and participants while maintaining compatibility with different language models.
  }
  \label{fig:teaser}
\end{teaserfigure}

\maketitle

\section{Introduction}

\ac{llm} and \ac{llm}-powered agents have transformed how analysts approach complex reasoning and information search tasks~\cite{Ferrag2025reasoning, DBLP:conf/acl/WuZLXJ25, DBLP:conf/bigdataconf/YousufDSXR24}.
Current systems operate through ephemeral prompts~\cite{DBLP:conf/nips/BrownMRSKDNSSAA20}: users specify requirements upfront, receive outputs, then manually refine results by correcting errors, reorganizing content, or adjusting terminology.
This workflow creates a one-way exchange: the system generates content, but user modifications never feed back to improve subsequent reasoning~\cite{DBLP:journals/frai/PasseriniGMST24, shridhar2023artllmrefinementask}.
When a domain expert edits an AI-generated artifact---correcting technical details, restructuring arguments, or refining specialized vocabulary---those modifications encode valuable domain expertise that remains latent and difficult to extract through explicit prompting alone~\cite{Patterson2010implicit, Miller2024context}.
Yet this knowledge does not propagate back into the system.
Instead, each new request triggers a cold start, forcing users to repeatedly re-communicate domain understanding that they have already demonstrated through their edits.

Consider a concrete scenario: a visualization researcher uses an \ac{llm}-based system to generate research questions from a recent paper.
The initial output contains three recurring error classes: 
(\textit{i}) nonsensical visualization suggestions (\textit{``use a 3D pie chart for temporal data''}), 
(\textit{ii}) technical inaccuracies (misusing terms such as ``semantic zoom'' when meaning ``geometric zoom''), and 
(\textit{iii}) wrong topical focus (emphasizing implementation details when the paper's contribution is conceptual).
The researcher corrects these errors, adjusting visualization types, fixing terminology, and redirecting focus to align with domain conventions.
However, when generating questions for the next paper, the system repeats similar mistakes, having learned nothing from the previous corrections.
This pattern forces experts to repeatedly correct the same fundamental misunderstandings about their domain.

The need for better human-AI collaboration is evident across domains.
In a preliminary survey with four pharmaceutical research professionals about their literature review workflows, we found that experts spend 5-60 hours on manual title/abstract screening, with teams processing 50-6,000 papers per review cycle.
Critically, 25-50\% of pre-processed datasets require manual corrections.
When asked about AI assistance, all participants found AI pre-labeling acceptable ``with spot-checks,'' but emphasized wanting to ``save time for crucial activities like categorization and extractions'' and ``deeper scientific analysis.''
Current AI systems require users to become prompting experts to effectively communicate domain requirements~\cite{DBLP:conf/chi/Zamfirescu-Pereira23,Mishra2025PromptAidVP}.
These findings highlight that domain experts need AI systems that can learn their preferences and domain expertise through natural interactions rather than explicit specification~\cite{Desmond2024ExploringPE,arawjo2023chainforge}.

To address this challenge, we pose three research questions:
\begin{itemize}
\item \textbf{RQ1:} How can user modifications to \ac{ai}-generated artifacts be systematically captured and transformed into reusable domain knowledge for multi-agent systems?
\item \textbf{RQ2:} What mechanisms enable bidirectional propagation of domain expertise between human edits and \ac{llm}-powered agent reasoning?
\item \textbf{RQ3:} How does accumulated implicit knowledge from multiple users improve subsequent artifact generation quality and reduce correction effort?
\end{itemize}

We introduce \textit{context-mediated domain adaptation}, a bidirectional human-\ac{ai} interaction~\cite{Amershi2019, Shneiderman2022} paradigm that treats user modifications as implicit domain specification capable of reshaping multi-agent reasoning behavior.
Building upon foundational work on context-mediated behavior in intelligent agents~\cite{turner1998context}, our approach extends these principles to modern \ac{llm}-powered multi-agent systems with persistent knowledge accumulation capabilities.
Our approach implements bidirectional semantic links through a structured artifact format enabling fine-grained editing and an adaptive context object representing accumulated domain knowledge.
This adaptive context is both trackable through comprehensive logging and systematically utilizable across sessions and participants.
When users modify artifacts through our three interaction modes, the system analyzes edit patterns to extract domain knowledge including terminology preferences, structural conventions, and conceptual relationships.
This extracted knowledge propagates back as enriched context through a formal adaptation mechanism involving edit distance calculation, prompt specificity analysis, and behavioral metrics tracking, enabling multi-agent systems to learn from human expertise.

We validate our approach through \textsc{Seedentia}, a web-based multi-agent framework enabling knowledge accumulation across participants.
An exploratory evaluation with five domain experts in visualization literacy demonstrates the feasibility of capturing implicit expertise through edit patterns, with 46 domain knowledge entries extracted from user modifications across research question generation tasks.

The contributions of this work are
(1) the \textbf{context-mediated domain adaptation (CMDA)} paradigm enabling bidirectional human-\ac{ai} interaction where user modifications reshape multi-agent behavior through structured knowledge representations (\ac{bdar} format and \ac{aco});
(2) a \textbf{prototype implementation} through Seedentia with three interaction modes (direct manipulation, prompt-based regeneration, context-based generation) and comprehensive logging infrastructure supporting cross-participant knowledge accumulation; and
(3) \textbf{exploratory evaluation results} with domain experts demonstrating the feasibility of capturing implicit expertise through edit patterns and cross-participant knowledge transfer.
 
\section{Related Work}

Our work builds upon four complementary research streams: context management in agentic systems, interactive machine learning systems that adapt through user feedback, prompt engineering frameworks that enable iterative refinement, and multi-agent architectures for complex reasoning tasks.
We position context-mediated domain adaptation as a synthesis that addresses limitations in each area while enabling bidirectional learning from user modifications.

\subsection{Multi-Agent Systems and Context Management}

Context management in agentic systems has evolved from early foundational work on context-mediated behavior~\cite{turner1998context} to sophisticated frameworks for explicit context representation and multi-agent collaboration.
Recent surveys highlight the importance of context acquisition, abstraction, and utilization pipelines in enabling agents to adapt and make robust decisions~\cite{du2024survey}.
Explicit context representation approaches~\cite{munguia2023deep,tutum2020generalization} demonstrate how separating context from skills enables agents to generalize to unseen situations, improving learning efficiency and robustness.
Multi-agent collaboration frameworks like Chain-of-Agents~\cite{zhang2024chain} and self-taught agentic systems~\cite{zhuang2025selftaught} show how agents can process long-context tasks through sequential communication and hierarchical reasoning.
However, these approaches focus primarily on task-specific context rather than accumulated domain expertise from user interactions.

\subsection{Human-AI Collaborative Knowledge Building}

The foundation for learning from user interactions was established by Endert et al.'s work on \textit{semantic interaction}, which demonstrated how user manipulations of visualizations can implicitly adjust underlying models, enabling domain expertise injection through intuitive interactions~\cite{endert2012semantic}.
This principle of implicit knowledge transfer through user actions directly informs our approach to extracting domain knowledge from edit patterns.

Mixed-initiative interaction~\cite{DBLP:journals/expert/Hearst99, DBLP:conf/chi/Horvitz99} systems---as early examples of human-centered AI~\cite{Shneiderman2022} and human-AI interaction~\cite{Amershi2019}---have long explored how systems can learn from user feedback, with early interactive machine learning work establishing principles that carry over to modern \ac{llm} systems.
However, these approaches typically focus on model training rather than real-time context adaptation within interactive sessions.

Recent work on AGDebugger~\cite{epperson2025debuggingmultiagent} demonstrates the importance of interactive message resets and editing capabilities for debugging multi-agent AI systems, validating our approach of bidirectional interaction.
Building trust in ML systems through visual explanations~\cite{yang2021visualexplanationstrustml} remains a critical challenge that our transparent edit tracking addresses.

While these systems enable user feedback integration, they lack mechanisms for persistent context evolution that spans multiple interaction cycles.
This is a gap that our bidirectional semantic links address.

\subsection{Prompt Engineering}

The emergence of sophisticated prompt engineering tools reveals the critical need for systematic approaches to \ac{llm} optimization~\cite{DBLP:conf/nips/BrownMRSKDNSSAA20}.
Prompting in general is difficult~\cite{DBLP:conf/chi/Zamfirescu-Pereira23}; non-experts lack good mental models, use opportunistic rather than systematic prompting techniques, and often overgeneralize their prompts.
As a result, examples abound in the human-centered AI~\cite{Shneiderman2022} and human-computer interaction disciplines of sophisticated prompting techniques hidden by interactive graphical user interfaces.
ChainForge~\cite{arawjo2023chainforge} provides visual toolkits for prompt engineering and hypothesis testing, demonstrating user demand for systematic prompt refinement capabilities.
PromptAid~\cite{Mishra2025PromptAidVP} offers visual prompt exploration and iteration capabilities, showing the value of systematic prompt development environments.

Enterprise applications have highlighted additional challenges in prompt engineering.
Desmond and Brachman~\cite{Desmond2024ExploringPE} identify key obstacles including the need for iterative refinement and domain-specific adaptation; precisely the problems our context-mediated approach addresses.
The ART framework~\cite{shridhar2023artllmrefinementask} introduces ask-refine-trust cycles for \ac{llm} improvement, establishing foundations for iterative refinement.
However, these approaches remain unidirectional: users refine prompts manually without systems learning from modification patterns to improve future interactions.

Research on dynamic system prompting shows how \ac{llm}s can adapt to real-time context changes, supporting our theoretical framework for context-mediated domain adaptation.
The PROMST framework~\cite{chen2024prompt} demonstrates the value of incorporating human feedback for prompt optimization in multi-step tasks, recognizing that humans excel at providing feedback about \ac{llm} outputs even when they struggle with direct prompt engineering---a principle that directly informs our bidirectional learning approach.
While prompt optimization work~\cite{zhou2023large} demonstrates that LLMs can generate effective prompts, and recent work shows LLMs can autonomously improve through implicit feedback~\cite{chen2024teaching}, these approaches require explicit specification of desired behaviors or operate at the model level rather than project-specific knowledge accumulation.
CMDA differs by capturing implicit domain knowledge through edit analysis, enabling systems to learn preferences users cannot easily articulate.
Yet existing prompt optimization frameworks lack the persistent memory and behavioral adaptation mechanisms that enable true domain specialization over time.

\subsection{Implicit Knowledge Extraction}

Current \ac{llm}-powered multi-agent frameworks excel at complex reasoning but lack mechanisms for incorporating user feedback into agent behavior modification.
Traditional sensemaking tools focus on information organization and visualization but do not leverage user interactions to improve underlying reasoning processes.
Work on human-AI collaboration~\cite{Amershi2019} emphasizes the importance of maintaining user agency while enabling system adaptation, principles central to our bidirectional interaction paradigm.
Recent participatory AI approaches~\cite{elmqvist2025participatoryaiscandinavianapproach} provide frameworks for meaningful human participation in AI system development and operation.

Modern conversational interfaces have begun addressing interaction persistence through memory systems.
ChatGPT and Claude now maintain ``memories''---explicit facts extracted from conversations through periodic introspection on dialogue history.
These systems scan transcripts for factual nuggets (user preferences, biographical details, stated constraints) and store them as retrievable context for future sessions.
While this represents progress beyond disposable prompts, memory extraction remains coarse-grained and declarative: systems capture what users explicitly state, not what they implicitly know.
Unlike vector database approaches that retrieve similar past examples or conversational memory systems that store explicit statements (e.g., ``User prefers Oxford comma''), CMDA extracts actionable operational patterns from user modifications (e.g., ``Research questions should specify target user expertise level'' inferred from edits that consistently add expertise qualifiers).
This enables behavioral adaptation rather than just contextual retrieval.

Recent work by Gao et al. on inferring latent user preferences from edit histories through LLM-based analysis demonstrates the viability of extracting implicit knowledge from user modifications, providing methodological foundations for our approach of learning domain expertise from artifact edits~\cite{DBLP:conf/nips/GaoTSMM24}.
However, their work focuses on personalizing outputs to individual stylistic preferences, whereas our approach extracts generalizable domain knowledge that transfers across users within a discipline.
While approaches like RLHF operate at token-level reward signals through gradient updates that optimize model weights, and transparency concerns~\cite{zhao2024explainability} motivate external knowledge mechanisms, CMDA operates at the semantic artifact level by enriching generation context without retraining.
This enables interpretable knowledge extraction (human-readable patterns), domain-specific adaptation (knowledge scoped to projects/users), and cross-participant transfer (one user's corrections improve others' generations).

Collaborative knowledge management research provides additional foundations for our approach.
Dörk et al.~\cite{dork2020codesigning} demonstrate how co-design processes that involve domain experts directly in visualization design ensure that resulting systems reflect specific practices, language, and needs of all users---principles that directly inform our context-mediated adaptation approach.
Peng et al.~\cite{peng2017collaborative} show how graph-based models can effectively combine formal knowledge with tacit expertise, enabling flexible, context-rich representations that support capturing and reusing knowledge as it evolves through collaboration.
Weck et al.~\cite{weck2021knowledge} explore knowledge management visualization in collaborative decision-making, demonstrating how visual representations can facilitate integration of multiple perspectives and domain expertise.

These collaborative knowledge approaches validate the importance of capturing tacit domain expertise through natural interactions, but they lack mechanisms for persistent context evolution that spans multiple interaction cycles.
Existing collaborative AI systems treat user input as external guidance rather than as a source of domain knowledge that can fundamentally reshape system behavior.
Multi-agent architectures typically employ fixed interaction patterns and lack the adaptive mechanisms necessary for context-mediated domain specialization.
Our work addresses these limitations by introducing bidirectional semantic links that enable multi-agent systems to evolve their reasoning patterns based on accumulated user modifications, creating a feedback loop that bridges human domain expertise with automated reasoning capabilities.

Recent work on agentic visualization~\cite{dhanoa2025agenticvisualization} provides a systematic framework for understanding autonomous and semi-autonomous components in visualization systems.
This framework identifies recurring design patterns that effectively balance computational agency with human control.
Building on established visualization techniques for interactive exploration~\cite{elmqvist2008rollingdice}, several systems exemplify these patterns: InsightsFeed~\cite{badam2017steering} implements progressive visual analytics with an insight timeline, DataSite~\cite{cui19datasite} employs proactive background computations, and Snowy~\cite{srinivasan21snowy} generates contextual utterance recommendations.
Recent \ac{llm}-based systems extend these patterns further: AVA~\cite{liu24ava} uses multimodal \ac{llm}s for autonomous visualization decisions, Data Formulator~\cite{dataformulator2024} transforms raw data into visualizations based on user-defined concepts, and InsightLens~\cite{insightlens2025} captures insights from conversational workflows.
Multi-agent approaches to visualization are emerging, with systems for automated visual data reporting~\cite{gyarmati2025composableagenticautomatedvisual} and narrative generation~\cite{wolter2025multiagentdatavisualizationnarrative} demonstrating coordinated agent workflows.

These systems demonstrate the value of agent role patterns (Forager, Analyst, Chart Creator, Storyteller), communication patterns (Insight Timeline, Progress Indicator, Provenance Log), and coordination patterns (Scouting, Swarming, Monitoring, Consolidating).
However, they lack mechanisms for bidirectional learning where user modifications reshape agent behavior---a gap our context-mediated adaptation addresses through persistent semantic links and behavioral adaptation mechanisms.

Our genuine novelty lies not in individual components (LLMs, multi-agent systems, edit tracking) but in their systematic integration: bidirectional semantic links that maintain provenance from edits through extracted knowledge to subsequent generation, structured knowledge representations (\ac{bdar} + \ac{aco}) enabling persistent cross-user accumulation, and multi-agent orchestration where knowledge extraction and generation phases operate in coordinated cycles.
This integration enables a new interaction paradigm where domain expertise flows bidirectionally between human modifications and AI reasoning, rather than remaining trapped in disposable prompt contexts.
 
\section{Context-Mediated Domain Adaptation}
\label{sec:framework}

Through a process we call \textit{context-mediated domain adaptation} (CMDA), \ac{llm}-powered reasoning systems can develop domain-specific behaviors by observing and learning from bidirectional interactions modifying generated artifacts (whether text, visualizations, or structured narratives).
This approach builds upon established context management principles in agentic systems~\cite{du2024survey,krishnan2025advancing}, recent advances in vision-language models for visualization understanding~\cite{gyarmati2025visionlanguagemodelsvisualizationslike}, agentic visualization design patterns~\cite{dhanoa2025agenticvisualization}, and collaborative knowledge management techniques~\cite{peng2017collaborative,neogy2020representing}.

Our framework transforms the traditional unidirectional prompt-response paradigm into a bidirectional learning system where user modifications serve as implicit domain specification.
When users edit system outputs---correcting terminology, restructuring content, or refining domain-specific conventions---these modifications encode valuable expertise that propagates back to improve subsequent reasoning.
This creates an evolutionary process where generic prompts bootstrap into sophisticated domain specifications through iterative refinement cycles, enabling a fundamentally different interaction paradigm where systems learn from how users modify outputs rather than just from what users request.

\subsection{Formal Definitions}

\begin{definition}[Context-Mediated Domain Adaptation]
\label{def:cmda}
Context-Mediated Domain Adaptation (CMDA) is a bidirectional learning process where:
(1) user modifications $M = \{m_1, ..., m_n\}$ to system-generated artifacts $A = \{a_1, ..., a_n\}$ are systematically captured,
(2) domain knowledge $D$ is extracted through the knowledge extraction pipeline $f: M \rightarrow D$, which analyzes edit patterns to identify domain-specific terminology, methodological preferences, and conceptual refinements, and
(3) domain knowledge propagates back to reshape agent behavior through the context injection mechanism $g: D \rightarrow C$, where $C$ represents the enriched context that is automatically incorporated into agent prompts for subsequent generations.
\end{definition}

\begin{definition}[Bidirectional Domain-Adaptive Representation]
\label{def:bdar}
A \ac{bdar} is a structured artifact format that maintains persistent links between AI-generated content and user modifications, enabling knowledge extraction through state comparison.
Each artifact preserves generation context (prompts, parameters) and modification history (edit distances, change patterns), creating a delta-based representation capturing the difference between system knowledge and user expertise.
Bidirectionality creates a learning cycle: user edits generate knowledge signals that are extracted, accumulated, and propagated back to enrich future generations, enabling subsequent users to benefit from previous corrections without explicit re-specification.
This creates a continuous improvement loop where each user interaction simultaneously contributes to and benefits from collective domain expertise.
\end{definition}

\begin{definition}[Adaptive Context Object]
\label{def:aco}
The Adaptive Context Object represents accumulated domain knowledge extracted from user modification patterns, organized into three primary categories:

\begin{itemize}
    \item \textbf{Domain Terminology Evolution}: Systematic vocabulary preferences and specialized language usage patterns derived from user corrections;
    \item \textbf{Methodological Refinements}: Improvements to research approaches, analytical frameworks, and domain-specific practices; and
    \item \textbf{Conceptual Depth Changes}: Theoretical nuances, conceptual relationships, and domain-specific considerations not captured in initial generations.
\end{itemize}

Knowledge entries maintain provenance links to their source interactions while supporting scoped application with user-specific knowledge taking precedence over project-shared and global contexts.
\end{definition}

\begin{table}[htb]
    \centering
    \caption{\textbf{Knowledge categories extracted from user edits.}
    The system identifies three distinct types of domain knowledge from user modifications to generated artifacts.
    Domain terminology evolution captures vocabulary preferences through direct text corrections.
    Methodological refinements encode expert knowledge about research practices and analytical frameworks.
    Conceptual depth changes add theoretical nuance and scholarly connections that distinguish expert from novice discourse.}
    \label{tab:aco_categories}
    \begin{tabular}{p{3.5cm}p{4.5cm}p{5cm}}
    \toprule
    \textbf{Knowledge Category} & \textbf{Description \& Characteristics} & \textbf{Example Knowledge Patterns} \\
    \toprule
    \textbf{Domain Terminology Evolution} & 
    Systematic vocabulary preferences and specialized language usage patterns derived from user corrections to technical terms and domain-specific expressions &
    Replacing ``chart'' with ``visualization,'' preferring ``participants'' over ``users,'' adopting field-specific acronyms and technical terminology \\
    \midrule
    \textbf{Methodological Refinements} & 
    Improvements to research approaches, analytical frameworks, and domain-specific practices reflecting expert knowledge of proper methodologies &
    Specifying statistical analysis requirements, adding ethical considerations, refining experimental design elements, emphasizing reproducibility standards \\
    \midrule
    \textbf{Conceptual Depth Changes} & 
    Theoretical nuances, conceptual relationships, and domain-specific considerations that add scholarly depth beyond surface-level content &
    Adding theoretical frameworks, clarifying causal relationships, introducing domain-specific constraints, connecting concepts to established literature \\
    \bottomrule
    \end{tabular}
\end{table}

Table~\ref{tab:aco_categories} illustrates how different types of user interactions contribute to distinct knowledge accumulation patterns.
Direct manipulation typically captures terminology preferences through immediate text corrections, while prompt-based regeneration reveals methodological knowledge through user guidance on content restructuring.
Context-based generation leverages all accumulated knowledge categories to produce artifacts that reflect learned domain conventions without explicit user specification.

\subsection{Bidirectional Semantic Links}
\label{sec:bidirectional-links}

The core mechanism enabling domain adaptation is the maintenance of bidirectional semantic links between generated artifacts and their creation context.
Each artifact maintains comprehensive metadata including original prompts, generation parameters, and contextual information.
When users modify these artifacts, the system captures not just the changes but also their semantic relationship to the generation context.

We define three interaction modes that contribute to domain learning, inspired by established Human–AI Interaction paradigms~\cite{vanberkel2021humanaiinteraction,gammelgard2024humanaiinteraction}.
Rather than modeling general interaction timing or initiative, our modes formalize how user edits become machine-readable learning signals that drive persistent domain adaptation across sessions.

\directmanipulation captures explicit user interactions such as data modification and selection, revealing fine-grained terminology preferences, structural conventions, and localized corrections through immediate inline edits. 
This mode provides high-resolution signals about domain expertise and supports focused refinements.

\promptbasedregeneration enables LLM completions guided by a user prompt and contextual information, allowing users to express higher-level conceptual goals and methodological requirements in natural language. 
By analyzing both the prompt and resulting changes, the system extracts domain-specific preferences about research framing, rigor, and evaluation strategies.

\contextbasedgeneration produces LLM completions based purely on accumulated implicit knowledge without further user interaction, demonstrating learned domain understanding. 
This mode operationalizes previously extracted insights to generate new artifacts, creating a continuous learning loop in which each interaction type contributes distinct knowledge patterns that feed into subsequent generation contexts and support ongoing domain adaptation without explicit specification.

\subsection{Knowledge Extraction and Propagation}
\label{sec:knowledge-extraction}

Our approach implements explicit context representation principles~\cite{munguia2023deep,tutum2020generalization} by separating accumulated domain knowledge from task-specific skills, enabling generalization across different content generation scenarios.
The system extracts domain knowledge from user modifications through pattern analysis across three primary categories that represent different levels of domain expertise:

\begin{itemize}
    \item \textbf{Domain Terminology Evolution}: Captures surface-level refinements in language and vocabulary preferences.
    By comparing original and modified text, the system identifies domain-specific terminology---for instance, changing ``data sources'' to ``data of diverse modalities'' or replacing broad task categories with specific patterns like ``lookup, search, filtering.''
    These refinements teach the system precise language that distinguishes expert discourse from generic descriptions.

    \item \textbf{Methodological Refinements}: Encode procedural knowledge about research approaches and assessment techniques.
    These modifications reveal how experts conceptualize research methodologies, such as emphasizing ``physiological signals alongside visual attention'' or incorporating ``real-time user state monitoring'' into assessment paradigms.
    The system learns preferred evaluation frameworks, experimental designs, and analytical approaches from these patterns.

    \item \textbf{Conceptual Depth Changes}: Represent the deepest level of expertise, fundamentally expanding the system's understanding of research scope and implications.
    These changes introduce new theoretical frameworks (e.g., ``cognitive load theory''), expand to include accessibility considerations (e.g., ``assist low-vision users''), or establish cross-domain connections.
    Such modifications teach the system to approach problems with greater sophistication and broader perspectives.

\end{itemize}

Following established context management pipelines~\cite{du2024survey}, this extracted knowledge propagates through the system via enriched context that influences subsequent agent reasoning.
The LangGraph workflow orchestrates this propagation through the three-stage process of context acquisition (edit tracking), abstraction (knowledge extraction), and utilization (context injection):
(1) analyzing edit patterns to extract domain signals,
(2) updating behavioral metrics to track adaptation progress, and
(3) injecting learned knowledge into agent prompts for future generations.

This extraction and propagation process enables three key theoretical properties of our framework:

\begin{itemize}
    \item \textit{Specification Bootstrapping} allows vague initial prompts to evolve into precise domain specifications through iterative human-AI collaboration, where users need not explicitly articulate all domain requirements upfront.
    \item \textit{Implicit Knowledge Transfer} enables domain expertise embedded in user edits to transfer to the system without explicit programming, allowing experts to share knowledge through natural interactions rather than formal specification.
    \item \textit{In-Context Learning} ensures agent behavior adapts dynamically within working memory based on observed correction patterns, unlike traditional fine-tuning approaches.
\end{itemize}

These properties transform every user edit from a one-time correction into a persistent learning signal that improves future system behavior.
 
\section{Implementation}

\textsc{Seedentia} is implemented as a working prototype demonstrating CMDA principles in a production-ready web-based multi-agent framework.
The implementation fully realizes the domain adaptation mechanisms described in our theoretical framework.

\subsection{System Architecture}

To enable effective context-mediated adaptation, we implement a modern web-based architecture that supports bidirectional learning and persistent knowledge accumulation.
Our system separates concerns across three primary layers that work together to capture, process, and utilize domain knowledge from user interactions:

\begin{itemize}
    \item \textbf{Presentation Layer}: Built using Next.js 15 with React components providing interactive interfaces for artifact editing with real-time feedback and multiple interaction modes.
    The \texttt{interactive-report-viewer} component enables inline content modification with entity-level granular control, allowing users to modify individual research questions, abstracts, and contextual narratives.
    User modifications trigger immediate UI updates while simultaneously capturing edit signals for knowledge extraction.

    \item \textbf{Processing Layer}: Orchestrated through Python FastAPI backend with LangGraph workflow engine managing multi-agent coordination.
    The system analyzes user modifications to extract domain knowledge, managing the bidirectional flow between user actions and system adaptation.
    Knowledge extraction nodes integrated into the LangGraph workflow analyze edit patterns in real-time, comparing initial and final values to identify implicit domain expertise.
    The unified state management system implements conditional routing between different task types while maintaining semantic links between user edits and generation context.

    \item\textbf{Persistence Layer}: Implemented through PostgreSQL database with specialized schemas for knowledge persistence and comprehensive edit tracking.
    This layer maintains the bidirectional semantic links essential for domain adaptation while enabling real-time adaptation based on user feedback across multiple participants and sessions.

\end{itemize}

\subsection{User Interface}

Our implementation provides comprehensive infrastructure for the interaction modes defined in our framework (Section~\ref{sec:bidirectional-links}).
The user interface implements the three interaction modes defined in our framework through a sophisticated component architecture that prioritizes usability while capturing meaningful interaction signals for domain knowledge extraction.

\begin{figure*}[htb]
  \centering
  \begin{subfigure}[t]{0.49\textwidth}
    \centering
    \includegraphics[width=\textwidth]{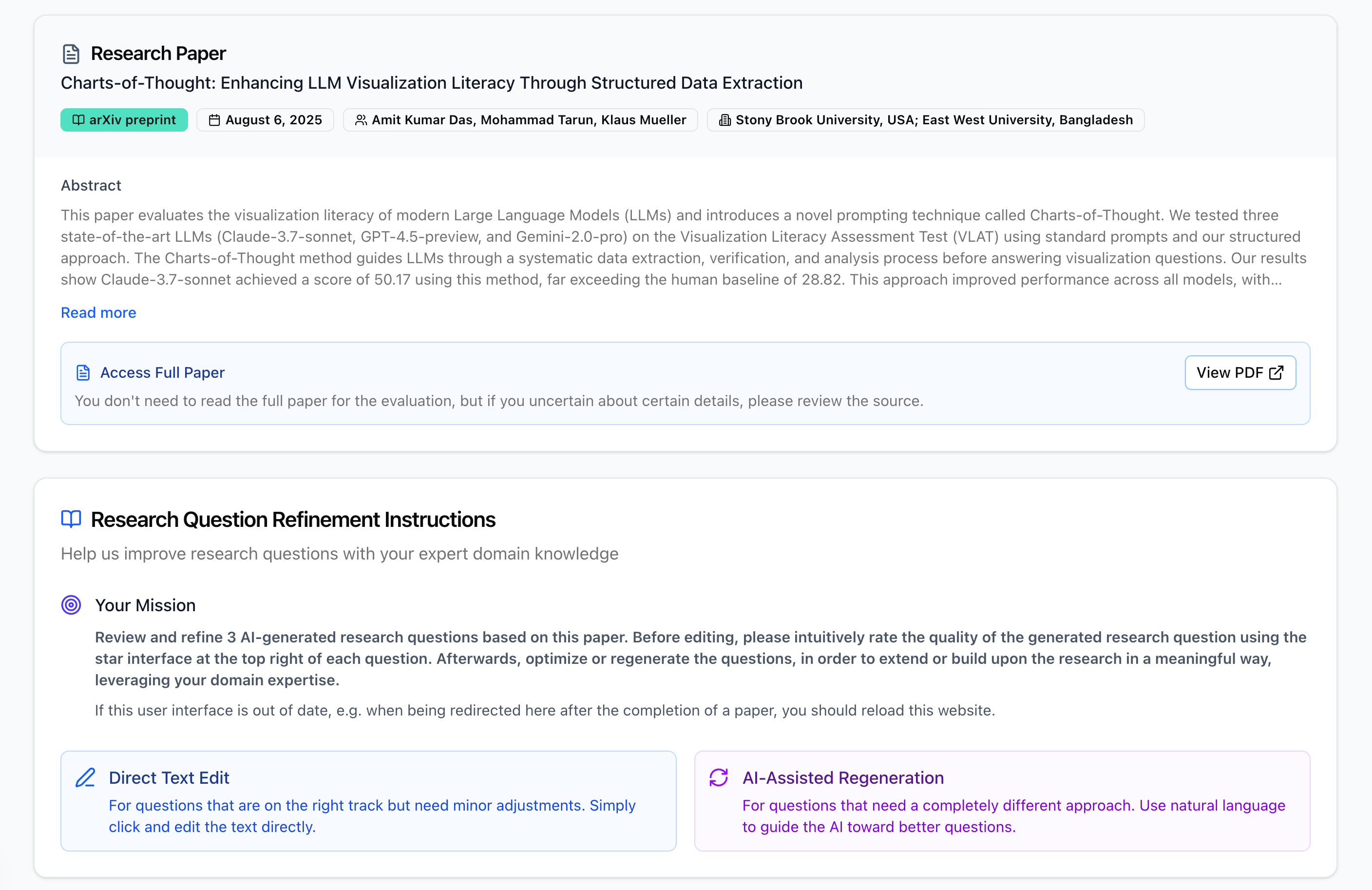}
    \caption{\textbf{Evaluation session initialization.}
    Interface showing the study setup where participants access paper details and interaction guidelines.
    This screen introduces domain experts to the research paper context and available interaction modalities before beginning the research question generation task that drives the context-mediated domain adaptation process.
    }
    \label{fig:evaluation_step_a}
  \end{subfigure}
  \hfill
  \begin{subfigure}[t]{0.49\textwidth}
    \centering
    \includegraphics[width=\textwidth]{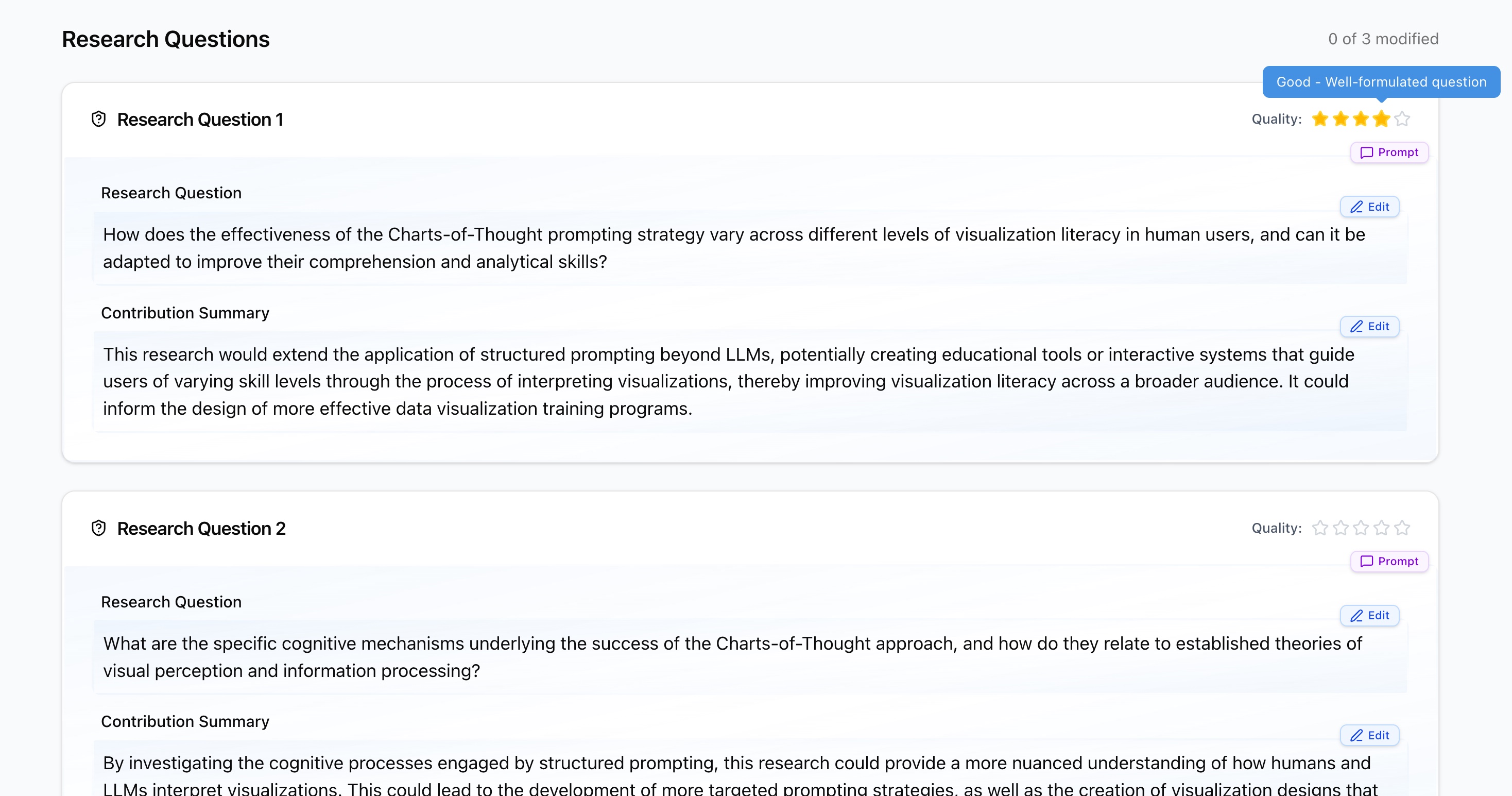}
    \caption{\textbf{Initial generation quality assessment.}
    Participant interface for rating AI-generated research questions on a 1-5 scale before editing begins.
    This baseline quality measurement enables quantification of context-mediated adaptation effectiveness by tracking improvement in initial generation quality as domain knowledge accumulates across participants, implementing our cross-user knowledge transfer evaluation metric.
    }
    \label{fig:evaluation_step_b}
  \end{subfigure}
  \caption{\textbf{Evaluation protocol interface.} 
  Key components of the controlled study we conducted for assessing context-mediated domain adaptation effectiveness. 
  The system captures baseline quality assessments and provides standardized interaction protocols to ensure consistent evaluation of bidirectional learning mechanisms across participants and sessions.}
  \label{fig:evaluation_starting}
  \Description{A side-by-side view of two evaluation protocol interfaces. The left panel (a) shows the evaluation session initialization screen with a header "Research Paper" displaying the title "Chart-of-Thought: Enhancing LLM Visualization Literacy Through Structured Data Extraction" with a green "FULLY ASSESSED" badge, author list, and affiliation. Below is an "Abstract" section with paper summary text, followed by an "Access Full Paper" section with a "View PDF" button. At the bottom is a "Research Question Refinement Instructions" box with purple header, containing "Your Mission" text explaining the participant's task and two instruction boxes: a blue "Direct Tool Edit" box stating "For questions that are not right track well (most major adjustments), freely clear out edit the text directly" and a pink "AI-Assisted Regeneration" box stating "For questions that aren't quite there (smaller approach or tone tweaks/range edits) guide the AI via prompt-specific questions." The right panel (b) shows the "Research Questions" interface with "1 of 3 Questions" header and a "Generate More" button. It displays "Research Question 1" with a 5-star quality rating, followed by the research question text about Chart-of-Thought prompting strategy and a "Contribution Summary" section. Below is "Research Question 2" marked as "DELETED" with similar structure showing a question about cognitive mechanisms and structured prompting.}
\end{figure*}

Users begin by accessing the paper details interface (Figure~\ref{fig:evaluation_step_a} from Section~\ref{sec:evaluation}), which displays metadata and provides a ``Generate Questions'' button to trigger initial context-based generation.
The system queries accumulated domain knowledge and generates three research questions asynchronously (10-30 seconds), each wrapped in an \texttt{AIContentWrapper} component exposing direct manipulation and prompt-based regeneration modes through color-coded interaction badges.

\subsubsection{Interaction Mode Architecture}

\begin{figure*}[th]
    \centering
    \begin{subfigure}[t]{0.49\textwidth}
      \centering
      \includegraphics[width=\textwidth]{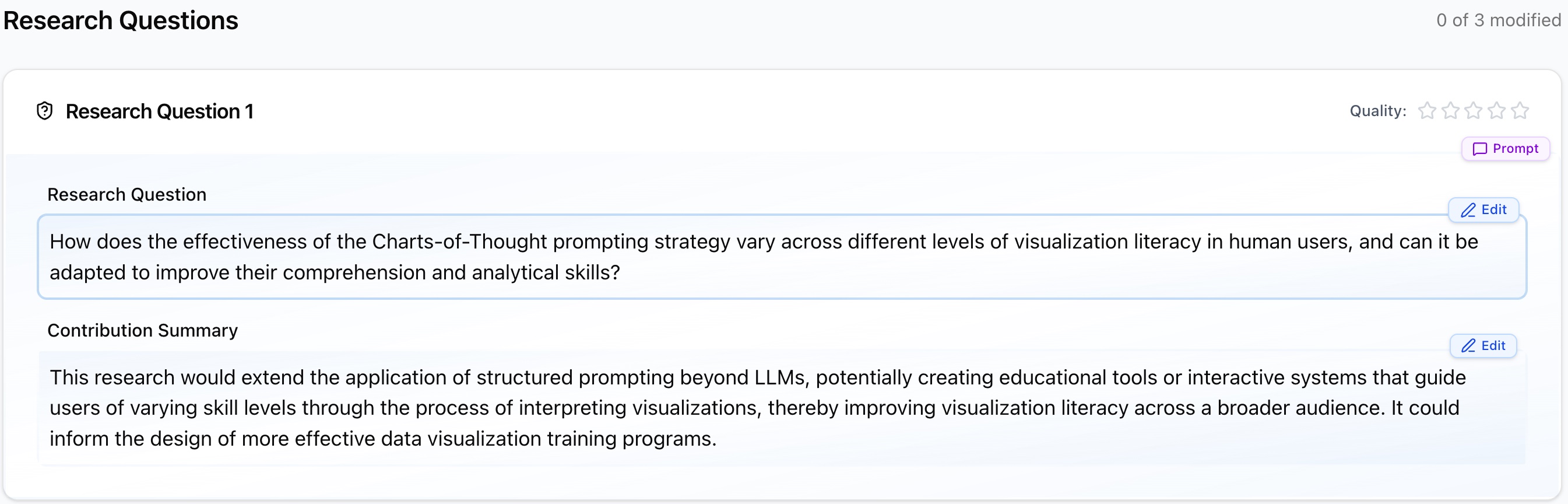}
      \caption{\textbf{Direct manipulation mode.}
        Interface showing direct content editing capabilities where users can modify research questions inline. 
        The border is highlighted on hover and indicates the editable content area, enabling immediate text modifications that are captured for domain knowledge extraction through the bidirectional semantic links described in our framework.
      }
      \label{fig:interaction_modality_direct}
    \end{subfigure}
    \hfill
    \begin{subfigure}[t]{0.49\textwidth}
      \centering
      \includegraphics[width=\textwidth]{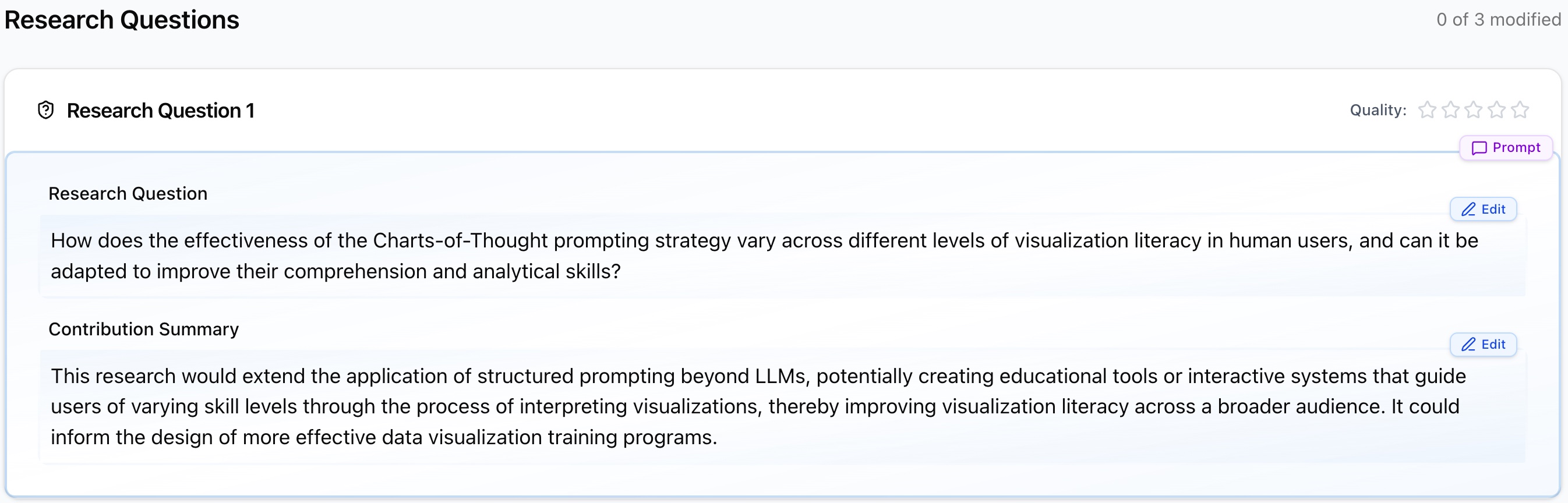}
      \caption{\textbf{Prompt-based regeneration mode.} 
      Interface demonstrating AI-assisted content regeneration where users provide natural language instructions to modify research questions.
      The border is highlighted on hover and indicates content that will be regenerated based on user prompts, implementing the context-mediated adaptation mechanism through explicit user guidance.
      }
      \label{fig:interaction_modality_prompt}
    \end{subfigure}
    \caption{\textbf{Interaction modalities.} 
    Implementation of interaction modes defined in our Context-Mediated Domain Adaptation framework.
    These interfaces demonstrate how user modifications are captured and transformed into domain knowledge through bidirectional semantic links, enabling the system to learn from expert corrections and improve subsequent artifact generation.}
    \label{fig:interaction_modalities}
    \Description{A side-by-side comparison of two interface screenshots. The left panel (a) shows "Direct Manipulation Mode" with a research question interface displaying "Research Question 1" at the top, followed by three sections: "Research Question" with editable text about Chart-of-Thought prompting strategy, "Contribution Summary" with a paragraph of text, and "Edit" buttons in the top right of each section. A red border highlights the editable content area. The right panel (b) shows "Prompt-based Regeneration Mode" with an identical layout, but includes a highlighted yellow section indicating content that will be regenerated based on user prompts. Both interfaces show quality indicators and word counts (25 words) in the top right corner. The screenshots demonstrate different interaction approaches for modifying AI-generated research questions.}
\end{figure*}

The system implements three distinct interaction modes.
Figure~\ref{fig:interaction_modalities} demonstrates the implementation of these interaction modes through our user interface design.
The interface prioritizes visual clarity and immediate feedback, with hover states that clearly indicate editable content areas and action boundaries.
Design decisions focus on minimizing cognitive load while maximizing the capture of meaningful interaction signals for domain knowledge extraction.
When users hover over editable content, the badges appear with a scale transform providing subtle visual feedback.
The implementation is purely based on CSS based hover pseudo classes without JavaScript overhead.
The system uses Tailwind's \texttt{peer} and \texttt{peer-hover} classes, enabling interaction badges to respond to hover states of their container elements.
Each interaction mode is color-coded according to the color coding within this paper itself:

\begin{itemize}
    \item[\textbf{blue}] for \textbf{\directmanipulation}: Enables inline content modifications through direct user input using the \texttt{useInlineEditor} hook.
    Hovering over generated content reveals the blue ``Edit'' badge; clicking transforms text into an editable field.
    For example, changing ``How can visualization systems assist users'' to ``How can interactive visualization systems assist low-vision users'' triggers three backend processes: 
    (1) database persistence with original value preserved, 
    (2) edit distance calculation, and 
    (3) knowledge extraction analyzing the added methodological qualifier and accessibility consideration.

    \item[\textbf{purple}] for \textbf{\promptbasedregeneration}: Facilitates AI-assisted content updates through natural language instructions via modal dialogs (Figure~\ref{fig:interaction_edit_prompt}).
    Clicking the purple ``Regenerate'' badge opens a dialog where users enter instructions (e.g., ``Make this question more specific to eye-tracking methodologies and real-time adaptation'').
    The targeted content displays a loading spinner during regeneration (5-15 seconds) while other interface elements remain fully interactive.
    Both the user's prompt and resulting modifications are analyzed to extract domain preferences (e.g., ``User values specific methodological details and real-time system characteristics'').

    \item[\textbf{amber}] for \textbf{\contextbasedgeneration}: Supports complete artifact creation based purely on accumulated context and previous user interactions.
    When triggered initially or for subsequent papers, the system queries the \texttt{implicit\_domain\_knowledge} table and automatically injects extracted terminology preferences, methodological refinements, and conceptual patterns into generation prompts.
    This creates a bidirectional learning cycle: user modifications extract knowledge that enriches future generation contexts without requiring explicit re-specification.

\end{itemize}

Figure~\ref{fig:interaction_modalities} shows the first two interaction modes, as context-based generation runs initially and ideally requires no user interaction.
Based on hovering certain elements triggering the said interactions, the affected elements of the interactions are highlighted.

The generic \texttt{AIContentWrapper} component manages its own edit states and interaction modes.
It is used to wrap all kinds of interaction modes separately.
The component follows a declarative pattern where interaction capabilities are enabled through props:

\begin{lstlisting}[language=JavaScript, caption=AIContentWrapper component usage example, label=lst:aicontentwrapper, basicstyle=\ttfamily\footnotesize, breaklines=true, breakatwhitespace=false]
<AIContentWrapper
  entityType="research_question"
  entityId={question.id}
  directEdit={{
    value: question.text,
    onSave: handleSave,
    validateValue: validateQuestionText
  }}
  onPromptEdit={() => setPromptDialogOpen(true)}
  onContextEdit={() => triggerContextGeneration()}
  onViewHistory={() => setHistoryDialogOpen(true)}
>
  {/* Editable content */}
  <ResearchQuestionDisplay question={question} />
</AIContentWrapper>
\end{lstlisting}

This pattern enables any artifact type to become editable by wrapping it with appropriate handlers, maintaining separation of concerns between presentation and interaction logic.

\subsubsection{Edit History and Provenance Tracking}

Beyond direct editing capabilities, the wrapper component provides comprehensive edit history visualization through the \texttt{EditHistoryDialog} component.
The system uses \texttt{react-diff-viewer} to display character-level differences between edit states, enabling users to review their modification patterns and understand how their changes contribute to domain knowledge extraction.

\begin{figure*}[htb]
    \centering
    \begin{subfigure}[t]{0.49\textwidth}
      \centering
      \includegraphics[width=\textwidth]{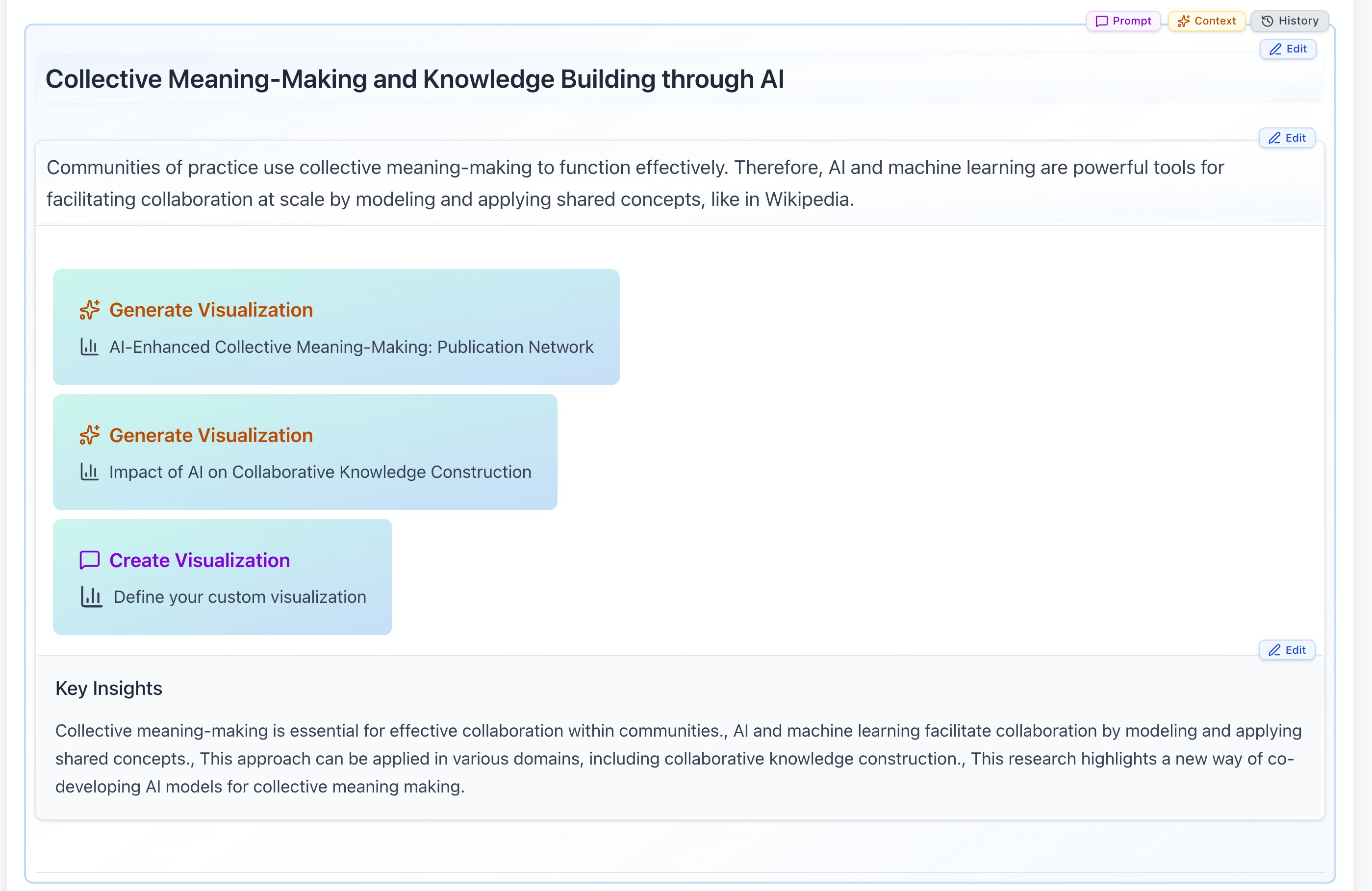}
      \caption{\textbf{AIContentWrapper with history access.}
        Interface demonstrating the \texttt{AIContentWrapper} component displaying multiple interaction mode badges alongside the edit history button (\textit{History} badge on the right).
        The component implements the CSS-based hover system where hovering over the interaction badges triggers the highlighting of the associated content area with a subtle border, providing immediate visual feedback about which artifact will be affected by user actions.
        This interface exemplifies how the wrapper component integrates seamlessly into complex nested content structures while maintaining clear interaction affordances.
      }
      \label{fig:interaction_ai_history_button}
    \end{subfigure}
    \hfill
    \begin{subfigure}[t]{0.49\textwidth}
      \centering
      \includegraphics[width=\textwidth]{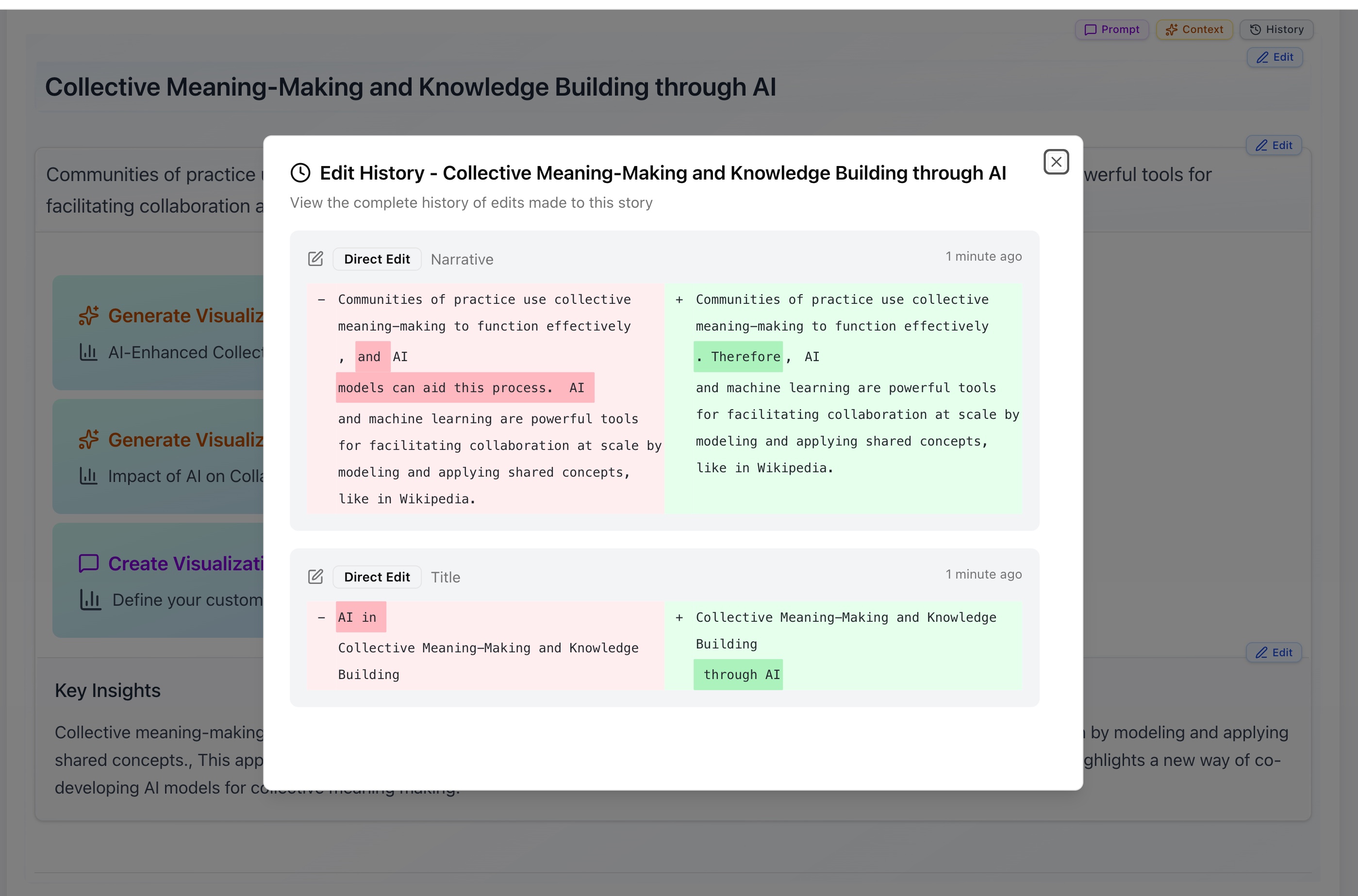}
      \caption{\textbf{Edit history dialog.} 
        Interface showing the comprehensive edit history dialog powered by \texttt{react-diff-viewer}, displaying character-level differences between edit states.
        The dialog presents a chronological timeline of modifications with metadata including edit timestamps, edit types (direct manipulation, prompt-based regeneration), and user context.
        Green highlights indicate additions while red highlights show deletions, enabling users to understand precisely how their interactions have modified the content over time.
        This granular tracking provides transparency into the bidirectional learning process and demonstrates how user modifications contribute to domain knowledge extraction.
      }
      \label{fig:interaction_ai_history_dialog}
    \end{subfigure}
    \caption{\textbf{Edit history visualization.} 
    The \texttt{AIContentWrapper} component provides integrated edit history functionality that powers context-mediated domain adaptation.}
    \label{fig:interaction_ai_history}
    \Description{A side-by-side comparison of edit history interfaces. The left panel (a) shows an AIContentWrapper component with a white background displaying the title "Collective Meaning-Making and Knowledge Building through AI" at the top, followed by three light blue boxes containing action items: "Generate Visualization" for Augmented Collective Meaning-Making Publication Network, "Generate Visualization" for Impact of AI on Collaborative Knowledge Construction, and "Create Visualization" to define a custom visualization. A "Key Insights" section appears below with explanatory text. In the top right corner is a "History" badge button with a clock icon. The right panel (b) shows a modal dialog titled "Edit History - Collective Meaning-Making and Knowledge Building through AI" overlaying a darkened version of the same interface. The dialog displays a chronological timeline with two entries labeled "Item edit - Interview" from 1 minute ago, each showing a diff view with green highlighted additions and red highlighted deletions in the text. The dialog includes tabs for "Interview" and "Title" at the top.}
\end{figure*}

Figure~\ref{fig:interaction_ai_history} demonstrates this functionality, showing both the history access button integrated into the wrapper component and the detailed diff view.
Each edit entry includes metadata such as edit type (direct edit, prompt modification, regeneration), timestamps, and provenance links to the generation context.
This transparency enables users to understand how their interactions shape the system's learning process while maintaining full visibility into the adaptation mechanisms.

\begin{figure*}[htb]
    \centering
    \begin{subfigure}[t]{0.49\textwidth}
      \centering
      \includegraphics[width=\textwidth]{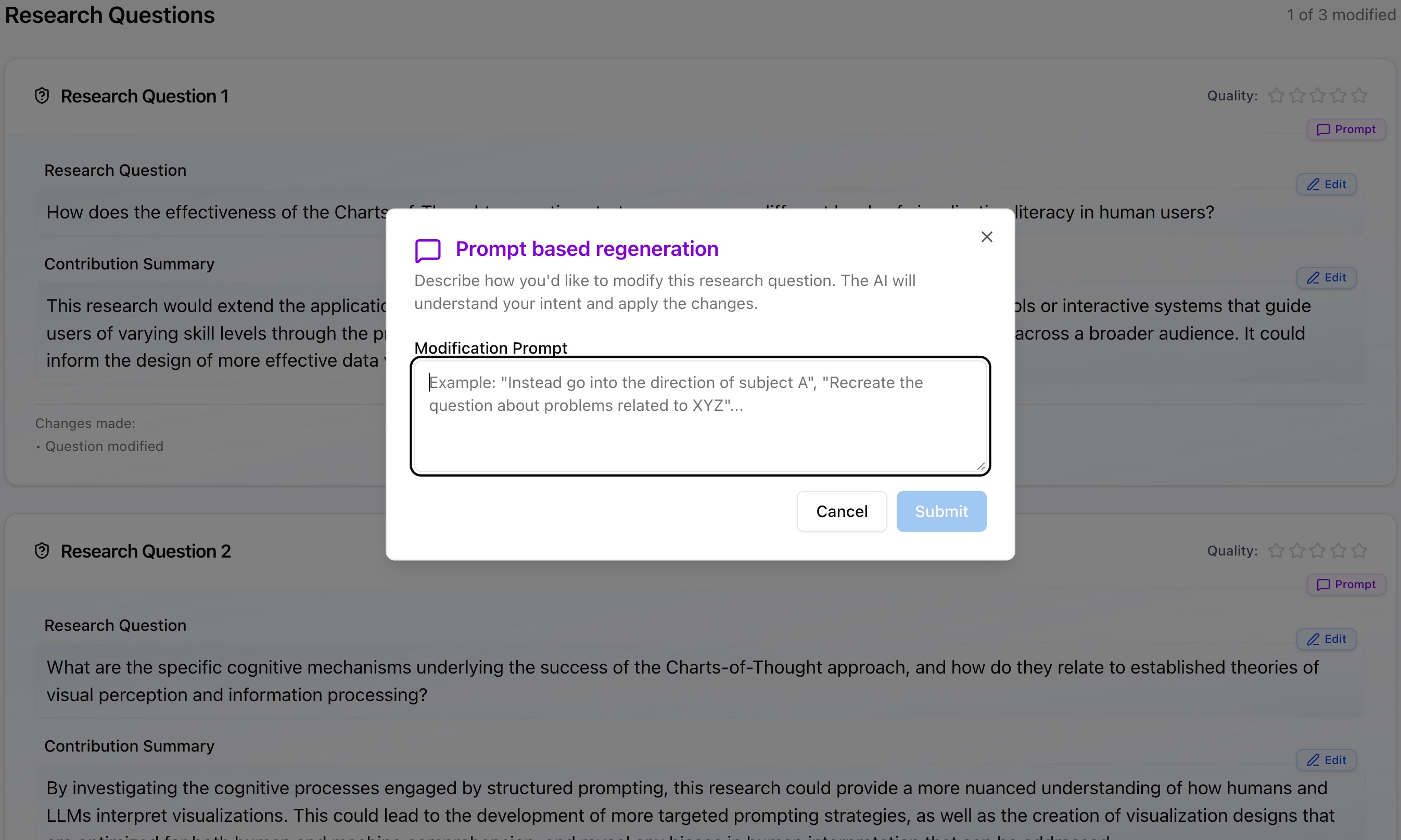}
      \caption{\textbf{Prompt-based input dialog.}
        Interface showing a dialog with a text area field for a user to insert custom prompt
      }
      \label{fig:interaction_edit_prompt_input}
    \end{subfigure}
    \hfill
    \begin{subfigure}[t]{0.49\textwidth}
      \centering
      \includegraphics[width=\textwidth]{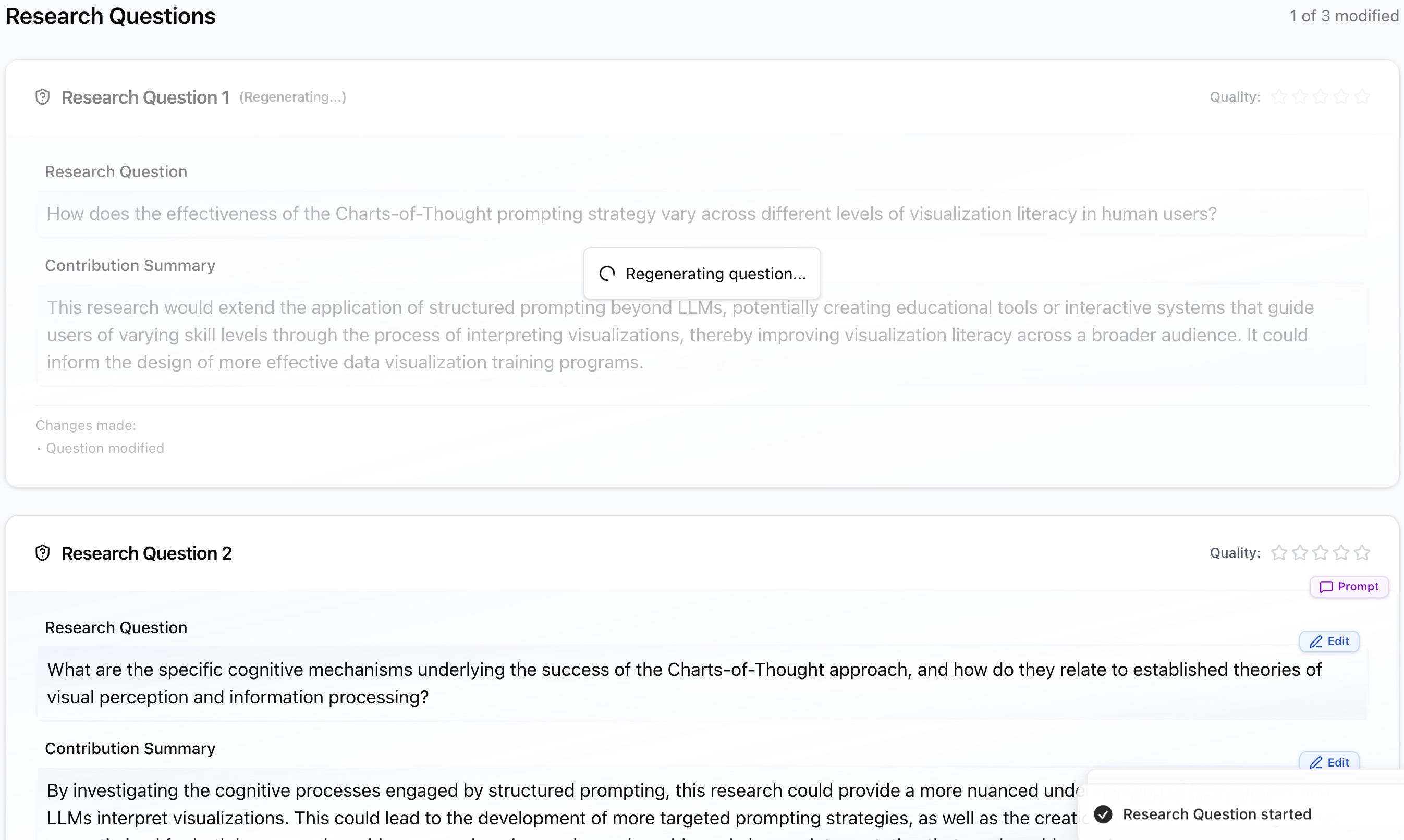}
      \caption{\textbf{Prompt-based regeneration.}
      Interface demonstrating a "loading state" showing that the respective artifact is being generated, while the rest of the application remains interactive
      }
      \label{fig:interaction_edit_prompt_generating}
    \end{subfigure}
    \caption{\textbf{Prompt-based generation.} 
    Complete workflow for prompt-based artifact regeneration showing the input dialog for natural language instructions and the asynchronous generation process.
    The interface maintains application responsiveness during AI processing, demonstrating the fire-and-forget architecture that decouples user interactions from computational workloads.
    }
    \label{fig:interaction_edit_prompt}
    \Description{A side-by-side comparison showing the prompt-based generation workflow. The left panel (a) displays a modal dialog box titled "Prompt based regeneration" overlaying a darkened research questions interface. The dialog contains explanatory text about how the system will regenerate content, a text input field labeled "Regeneration Prompt" with example prompts, and two buttons at the bottom: "Cancel" (gray) and "Regenerate" (blue). The right panel (b) shows the same interface during regeneration, with Research Question 1 displaying a spinner icon and grayed-out text indicating it is being regenerated, while Research Question 2 below remains fully visible and interactive with normal contrast. Both panels show the "Research Questions" header with "1 of 2 modified" status in the top right.}
\end{figure*}

When using prompt-based regeneration, the user can insert a prompt to regenerate the selected artifact.
Figure~\ref{fig:interaction_edit_prompt} shows the complete interaction flow from input dialog to asynchronous generation.

\subsubsection{Real-time State Management and Persistence}

The system implements optimistic UI updates through React state management, providing immediate feedback while asynchronous save operations complete in the background.
The Next.js application functions as both a sophisticated frontend and a lightweight server that creates tasks for the Python backend to execute.

Architecturally, the frontend operates as a standalone CRUD application optimized for high usability and user experience.
All complex AI processing, knowledge extraction, and multi-agent coordination is offloaded to the more sophisticated Python/LangGraph backend system.
This separation enables the frontend to remain responsive and maintain excellent user experience while computationally intensive operations execute asynchronously in the background.

The frontend communicates with the backend exclusively through the \texttt{agent\_tasks} infrastructure, using the \texttt{createAndTriggerAgentTask} pattern to initiate AI processing workflows.
Toast notifications provide non-intrusive feedback on save status and task completion, while maintaining focus on the editing experience.

\subsection{Core Infrastructure Implementation}

The user interface layer operates on a sophisticated infrastructure that manages data persistence, workflow coordination, and knowledge processing.
The core infrastructure implements CMDA through three integrated subsystems: database schema architecture, agentic task processing, and evaluation integration.

Our database schema implements \ac{bdar} (Definition~\ref{def:bdar}) through two distinct model categories.
Our architecture separates agentic task processing infrastructure from knowledge extraction mechanisms, enabling scalable multi-agent workflows while maintaining clean domain adaptation processing.

The separation between business domain models and agentic process models provides several key advantages:
(1) \textit{Frontend-Backend Decoupling}: The Next.js frontend communicates with the Python/LangGraph backend exclusively through the \texttt{agent\_tasks} infrastructure, eliminating the need for custom API endpoints.
(2) \textit{Asynchronous Processing}: Long-running AI operations execute without blocking the user interface, with real-time progress updates through Supabase subscriptions.
(3) \textit{Scalability}: The task queue system enables horizontal scaling of AI processing while maintaining clear separation between user-facing functionality and computational workloads.

\subsection{Agentic Task Processing Infrastructure}

The agentic task processing infrastructure forms the computational backbone of our Context-Mediated Domain Adaptation system, enabling scalable multi-agent workflows that extract and apply domain knowledge through sophisticated coordination mechanisms.

\textbf{Agentic System Process Entities} manage the asynchronous AI processing infrastructure that enables scalable multi-agent workflows:

\begin{table}[htb]
    \centering
    \caption{\textbf{Database tables supporting asynchronous multi-agent task processing.}
    The system uses six interconnected tables to coordinate AI agent workflows, track execution status, and log processing details.
    The \texttt{agent\_tasks} table serves as the central coordinator, while supporting tables define workflow types, decompose tasks into atomic actions, categorize processing steps, record external API calls, and maintain real-time execution logs for debugging and monitoring.}
    \label{tab:agentic_tables}
    \begin{tabular}{lp{8cm}}
    \toprule
    \textbf{Table} & \textbf{Purpose} \\
    \toprule
    \texttt{agent\_tasks} & Central coordination for asynchronous AI processing with status tracking via \texttt{status}, \texttt{input\_data}, and \texttt{output\_data} \\
    \midrule
    \texttt{task\_type} & Defines available AI workflows (research question generation, knowledge extraction, context injection) using \texttt{code} and \texttt{label} \\
    \midrule
    \texttt{task\_action} & Individual processing steps within multi-agent workflows for debugging using \texttt{attempts} and \texttt{error\_message} \\
    \midrule
    \texttt{action\_type} & Categorizes processing actions (extraction, analysis, generation) for workflow orchestration \\
    \midrule
    \texttt{api\_logs} & External API interactions for literature retrieval tracking \texttt{search\_terms} and \texttt{papers\_found} \\
    \midrule
    \texttt{project\_agent\_log} & Real-time logging of agent processing steps using \texttt{log\_type} and \texttt{message} for monitoring and debugging \\
    \bottomrule
    \end{tabular}
\end{table}

\subsubsection{Multi-Agent Coordination Pipeline}

\begin{figure*}[htb]
    \centering
    \includegraphics[width=0.80\linewidth, alt={Agentic task processing graph displays the langgraph nodes in a visualization.}]{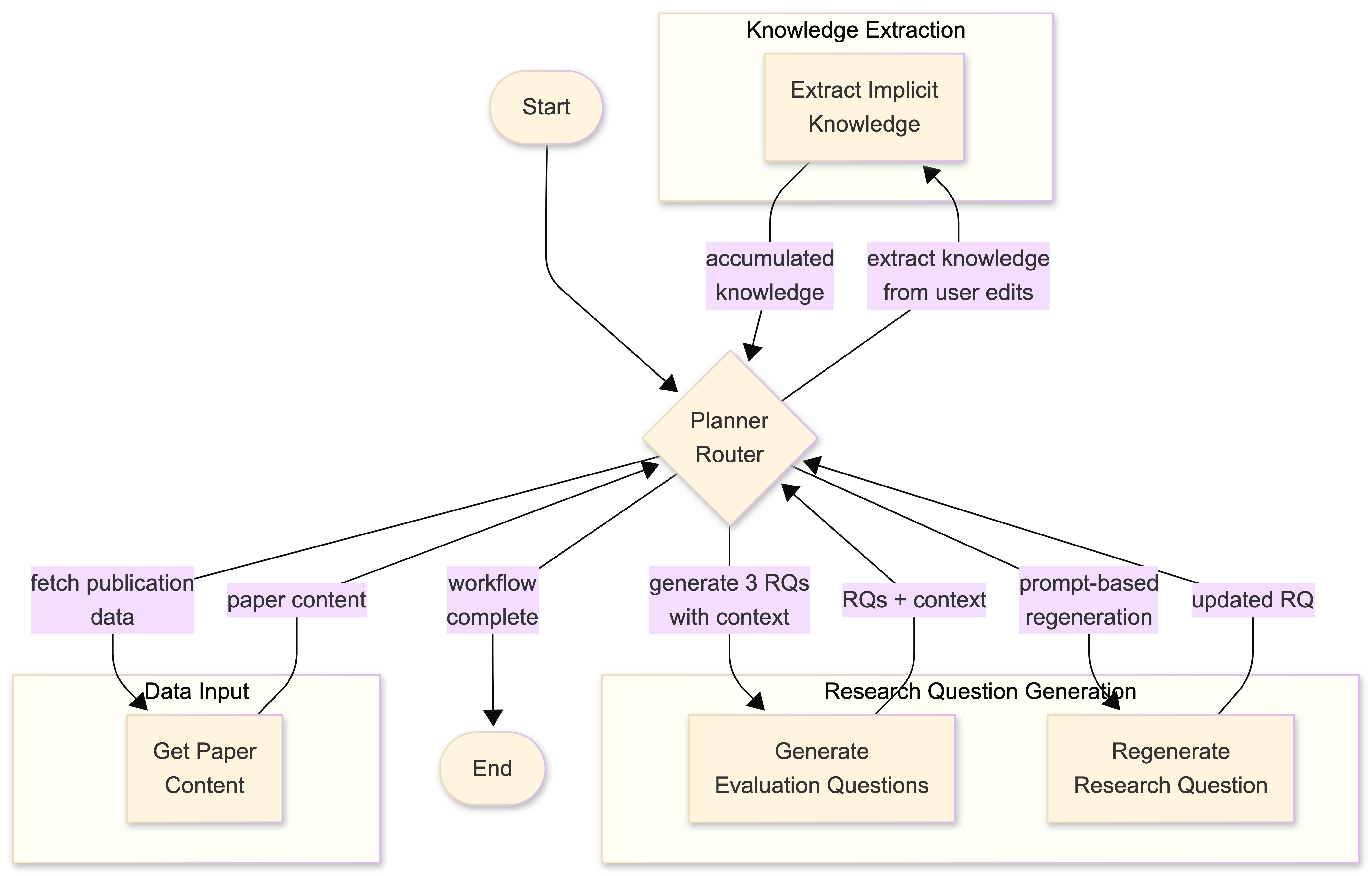}
    \caption{\textbf{Agentic task processing graph.} 
    The backend workflow graph is centered on the \texttt{planner} router node, which conditionally dispatches tasks to specialized nodes for paper retrieval, context-based research question generation, and edit-driven knowledge extraction.
    Node outputs are merged back into a unified state and persisted via the agent tasks infrastructure, enabling asynchronous execution while maintaining traceable bidirectional links between user edits, extracted domain insights, and subsequent generations.}
    \label{fig:seedentia_langgraph}
    \Description{
    Visualization of the backend workflow graph centered on the \texttt{planner} router node, which conditionally dispatches tasks to specialized nodes for paper retrieval (fetch\_paper\_content), context-based research question generation (generate\_evaluation\_questions), and edit-driven knowledge extraction (extract\_implicit\_knowledge).
    Node outputs are merged back into a unified state and persisted via the agent\_tasks infrastructure, enabling asynchronous execution while maintaining traceable bidirectional links between user edits, extracted domain insights, and subsequent generations.}
\end{figure*}

The LangGraph implementation provides sophisticated agent coordination through a unified planner/router that determines workflow paths based on task type and current state (Figure~\ref{fig:seedentia_langgraph}).
Following established patterns in multi-agent collaboration~\cite{zhang2024chain,krishnan2025advancing}, the system supports multiple specialized task types including research question generation, knowledge extraction, and context injection, with dynamic routing enabling context-aware workflow adaptation.
Key workflow capabilities include:

\begin{itemize}
    \item \textbf{State Consistency}: Unified state management maintains context across all workflow nodes, similar to Chain-of-Agents approaches for long-context processing
    \item \textbf{Interactive Control}: Pause/resume capabilities with user interrupt handling for real-time feedback integration
    \item \textbf{Node Modularity}: Specialized business logic separated into domain-specific nodes for maintainability and scalability
    \item \textbf{Context Propagation}: Seamless transfer of accumulated domain knowledge between processing stages
\end{itemize}

The workflow engine orchestrates the bidirectional flow described in our framework, implementing the context injection $g: D \rightarrow C$ mechanism (Definition~\ref{def:cmda}) by routing user modifications through specialized knowledge extraction nodes that analyze edit patterns and update persistent context repositories, which then influence future agent reasoning through enriched context.

\subsubsection{Asynchronous Processing Architecture}

The system implements a fire-and-forget processing pattern where the frontend creates tasks by inserting entries into the \texttt{agent\_tasks} table and receives UUIDs for tracking, while the Python backend asynchronously executes these tasks through the LangGraph workflow engine.
This architecture decouples user interface responsiveness from computational workloads, enabling complex knowledge processing without blocking user interactions.

Task execution follows a hierarchical structure where high-level tasks represent LangGraph workflow nodes and lower-level actions track individual processing steps.
The \texttt{task\_action} table records detailed execution context including attempts and error messages, enabling comprehensive debugging and progress tracking of multi-step operations.
External API interactions for literature retrieval are logged through \texttt{api\_logs}, maintaining search terms, result counts, and response metadata for reproducibility and system optimization.

\subsection{Domain Adaptation Engine}

The implemented system operationalizes the knowledge extraction and propagation mechanisms described in our framework through automated processing that forms the core of our context-mediated adaptation approach.

\subsubsection{Evaluation Knowledge Retrieval Model}

\textbf{Business and Functional Domain Models} represent the core domain concepts for context-mediated domain adaptation:

\begin{table}[htb]
    \centering
    \caption{\textbf{Database tables for knowledge extraction and evaluation tracking.}
    Eight tables implement the core knowledge representation infrastructure for context-mediated domain adaptation.
    The \texttt{evaluation\_research\_questions} table stores initial and final artifact states to enable knowledge extraction from user modifications.
    The \texttt{implicit\_domain\_knowledge} table materializes extracted domain expertise with category labels and provenance tracking.
    Supporting tables capture granular edit operations, preserve generation context for bidirectional semantic links, manage evaluation sessions with comprehensive metrics, profile participant expertise, store research papers for AI processing, and log detailed UI interactions for behavioral analysis.}
    \label{tab:business_tables}
    \begin{tabular}{lp{8cm}}
    \toprule
    \textbf{Table} & \textbf{Purpose} \\
    \toprule
    \texttt{evaluation\_research\_questions} & Implements \ac{bdar} by storing initial/final states for knowledge extraction using \texttt{initial\_question}, \texttt{current\_question}, and edit distance metrics \\
    \midrule
    \texttt{implicit\_domain\_knowledge} & Materializes \ac{aco} with categorized domain insights and provenance tracking through \texttt{knowledge\_category} and \texttt{source\_question\_ids} \\
    \midrule
    \texttt{ai\_entity\_edits} & Granular interaction tracking for behavioral analysis capturing \texttt{edit\_type}, \texttt{original\_value}, and \texttt{user\_prompt} \\
    \midrule
    \texttt{ai\_entity\_metadata} & Generation context preservation for bidirectional semantic links using \texttt{generation\_prompt} and \texttt{model\_parameters} \\
    \midrule
    \texttt{evaluation\_sessions} & Session management with comprehensive metrics including \texttt{edit\_distance\_score} and LLM monitoring via \texttt{langfuse\_trace\_id} \\
    \midrule
    \texttt{evaluation\_participants} & Participant management with \texttt{domain\_expertise} profiling and \texttt{evaluation\_status} tracking \\
    \midrule
    \texttt{publication\_raw} & Research paper storage with \texttt{full\_text} content for AI processing and evaluation study materials \\
    \midrule
    \texttt{user\_interactions} & Granular UI interaction tracking with \texttt{interaction\_type} and state transition analysis \\
    \bottomrule
    \end{tabular}
\end{table}

\subsubsection{Knowledge Extraction Pipeline}

The knowledge extraction pipeline implements the pattern analysis function $f: M \rightarrow D$ (Definition~\ref{def:cmda}) by processing all unprocessed user edits when research question generation tasks are initiated.
The pipeline analyzes differences between initial AI-generated content and final user-approved versions, categorizing insights into the three knowledge categories defined in our Adaptive Context Object (Definition~\ref{def:aco}):

\begin{itemize}
    \item \textbf{Domain Terminology Evolution} captures changes in specialized vocabulary, identifying when users consistently replace general terms with domain-specific language or prefer certain terminological conventions over others.

    \item \textbf{Methodological Refinements} identifies improvements to research approaches, study design considerations, or analytical frameworks that users introduce through their modifications.

    \item \textbf{Conceptual Depth Changes} encompasses modifications that add theoretical nuance, clarify conceptual relationships, or introduce domain-specific considerations not present in the initial generation.

\end{itemize}

Intelligent de-duplication ensures that similar insights are consolidated rather than duplicated across multiple extraction cycles, with each knowledge entry maintaining provenance links to its source interactions.

\subsubsection{Context Accumulation and Adaptive Generation}

\textbf{Context Accumulation}: Extracted knowledge accumulates across participants and sessions, creating a growing repository of domain expertise.
Each new generation benefits from previously extracted insights, enabling progressive improvement in output quality.
The system maintains provenance tracking, linking each piece of knowledge back to its source interactions while preventing redundant processing.

\textbf{Adaptive Generation}: When generating new artifacts, the system incorporates accumulated domain knowledge into the generation context.
This enables cross-participant learning where subsequent users benefit from the collective expertise of previous participants.
The knowledge injection occurs transparently, improving output quality without requiring users to explicitly specify domain requirements.

The research question generation process implements a sophisticated adaptive mechanism that evolves based on accumulated knowledge from participant interactions.
The \texttt{generate\_evaluation\_questions} function orchestrates this process by retrieving evaluation paper content, gathering accumulated knowledge from previous participants, and incorporating both explicit edits and implicit domain knowledge into the generation context.
This multi-layered knowledge integration ensures that each subsequent generation benefits from the collective expertise of all previous participants, creating a continuous improvement cycle.
The following prompt template is used for generating research questions that build upon published papers:

\begin{lstlisting}[caption=Research question generation prompt with knowledge accumulation., label=lst:research-questions-prompt, basicstyle=\ttfamily\footnotesize, breaklines=true, breakatwhitespace=false]
You are an expert researcher in Visualization Literacy who generates 
insightful research questions based on published papers. Your task is 
to generate exactly 3 research questions that extend or build upon the 
provided paper.

PAPER DETAILS:
Title: [paper_title]

Abstract: [paper_abstract]

Full Text: [paper_full_text]

GUIDELINES FOR RESEARCH QUESTIONS:
1. Each question should identify a research gap in the field
2. Each question should be agnostic and not require knowledge of the 
   paper itself to understand the question
3. Questions should be specific, measurable, and feasible for future 
   research

[If participant_order > 1:]
ACCUMULATED KNOWLEDGE FROM PREVIOUS PARTICIPANTS:
[knowledge_context including expansion, refinement, and condensation 
patterns extracted from previous participants' edits]

Based on the patterns above, generate questions that reflect these 
improvements and refinements. Learn from how previous participants 
enhanced their initial responses.

RESPONSE FORMAT:
Generate exactly 3 research questions. For each question, provide:
1. A clear, specific research question
2. A brief summary (2-3 sentences) of the potential contribution if 
   this research were conducted

Focus on questions that:
- Address limitations or gaps in the current paper
- Extend the methodology to new contexts or populations
- Explore long-term implications or applications
- Investigate underlying mechanisms or theoretical foundations
- Consider interdisciplinary connections
- Address scalability, generalizability, or practical implementation

Generate innovative, thought-provoking questions that a domain expert 
in Visualization Literacy would find valuable and feasible for future 
research.
\end{lstlisting}

When accumulated domain knowledge is available from previous participants, the system automatically injects this knowledge into the generation context, enabling progressive refinement of research question quality across evaluation sessions.

The multi-agent workflow orchestrates the complete adaptation cycle: extracting implicit knowledge from user modifications, categorizing and storing domain insights, and enriching future generations with accumulated expertise.
This creates a continuous learning loop where each interaction strengthens the system's domain understanding.
Crucially, every step of this process is comprehensively monitored and traced through integrated observability infrastructure, enabling both real-time operational monitoring and systematic research analysis of adaptation effectiveness.

\subsection{Evaluation and Knowledge Extraction Integration}

The evaluation and knowledge extraction integration provides comprehensive infrastructure for systematic assessment of domain adaptation effectiveness while operationalizing the knowledge extraction mechanisms described in our theoretical framework.

\subsubsection{Evaluation Framework Architecture}

Our implementation includes a sophisticated evaluation framework designed for controlled studies of domain adaptation effectiveness.
The system supports participant management with domain expertise profiling through the \texttt{evaluation\_participants} table, session-based tracking via \texttt{evaluation\_sessions}, and comprehensive interaction logging through \texttt{user\_interactions}.

The framework automatically calculates edit distances between initial AI-generated content and final user-approved versions at both character and word levels.
Edit distance computation occurs separately for research questions and contribution summaries, enabling fine-grained analysis of user modification patterns.
Time-to-completion metrics are tracked per session, while edit type frequencies and prompt evolution patterns provide insights into user behavior and system adaptation effectiveness.

\subsubsection{Knowledge Extraction Pipeline Integration}

The \texttt{evaluation\_research\_questions} table implements the core \ac{bdar} concept by storing original AI-generated content and final user-approved versions, enabling systematic knowledge extraction through comparison of initial and final states.
Each entry contains initial and final text states with calculated edit distances, providing self-contained units for LLM-based knowledge generation essential to our CMDA framework.

The knowledge extraction process employs a specialized prompt structure that analyzes the differences between original AI-generated content and final user-approved versions.
The following listing shows the prompt template used for extracting implicit domain knowledge:

\begin{lstlisting}[caption=Knowledge extraction prompt template., label=lst:knowledge-extraction-prompt, basicstyle=\ttfamily\footnotesize, breaklines=true, breakatwhitespace=false]
You are an expert in Visualization Literacy research who analyzes how 
domain experts refine AI-generated research questions. Your task is to 
extract implicit domain knowledge from the changes a user made.

ORIGINAL AI-GENERATED CONTENT:
Question: [initial_question]
Contribution: [initial_contribution]

FINAL USER-EDITED CONTENT:
Question: [final_question]
Contribution: [final_contribution]

[existing_knowledge if available]

ANALYSIS TASK:
Analyze the changes from original to final versions and extract 
implicit domain knowledge that reflects:

1. **Domain Terminology Evolution**: How the user refined technical 
   terms, concepts, or field-specific language
2. **Methodological Refinements**: Changes to research methods, 
   approaches, or evaluation criteria  
3. **Conceptual Depth Changes**: Shifts in research focus, scope, 
   specificity, or theoretical framing

EXTRACTION RULES:
- Only extract knowledge if there are meaningful changes 
  (not just minor rewording)
- Focus on domain expertise that could improve future AI-generated 
  questions
- Each insight should be actionable for improving AI generation
- Avoid extracting knowledge that duplicates existing entries
- Generate 0-3 knowledge entries based on the significance of changes

RESPONSE FORMAT:
Return a JSON array of knowledge objects. Each object must have:
- "text": Clear, actionable insight (1-2 sentences)
- "category": One of "domain_terminology_evolution", 
  "methodological_refinements", "conceptual_depth_changes"

Example format:
[
  {
    "text": "Research questions should specify the target population 
             (e.g., 'novice users' vs 'domain experts') rather than 
             using generic terms like 'users'.",
    "category": "domain_terminology_evolution"
  },
  {
    "text": "Evaluation studies in visualization literacy should 
             include both immediate comprehension and retention 
             measures over time.",
    "category": "methodological_refinements"
  }
]

If no meaningful domain insights can be extracted, return an empty 
array: []
\end{lstlisting}

The \texttt{implicit\_domain\_knowledge} table materializes the Adaptive Context Object by storing extracted insights categorized into the three primary knowledge categories: domain terminology evolution, methodological refinements, and conceptual depth changes.
Direct references to source research questions through \texttt{source\_question\_ids} arrays maintain traceable knowledge provenance, enabling detailed analysis of how specific user modifications contribute to accumulated domain expertise.

\subsubsection{System Monitoring and Observability}

Our implementation integrates comprehensive monitoring infrastructure to provide visibility into the context-mediated adaptation process.
The monitoring architecture combines Langfuse~\cite{langfuse} for \ac{llm} interaction tracing with custom instrumentation for tracking domain adaptation effectiveness, creating a multi-layered observability system that supports both operational reliability and systematic research validation.

\begin{figure*}[htb]
    \centering
    \includegraphics[width=\linewidth, alt={Seedentia LLM tracing.}]{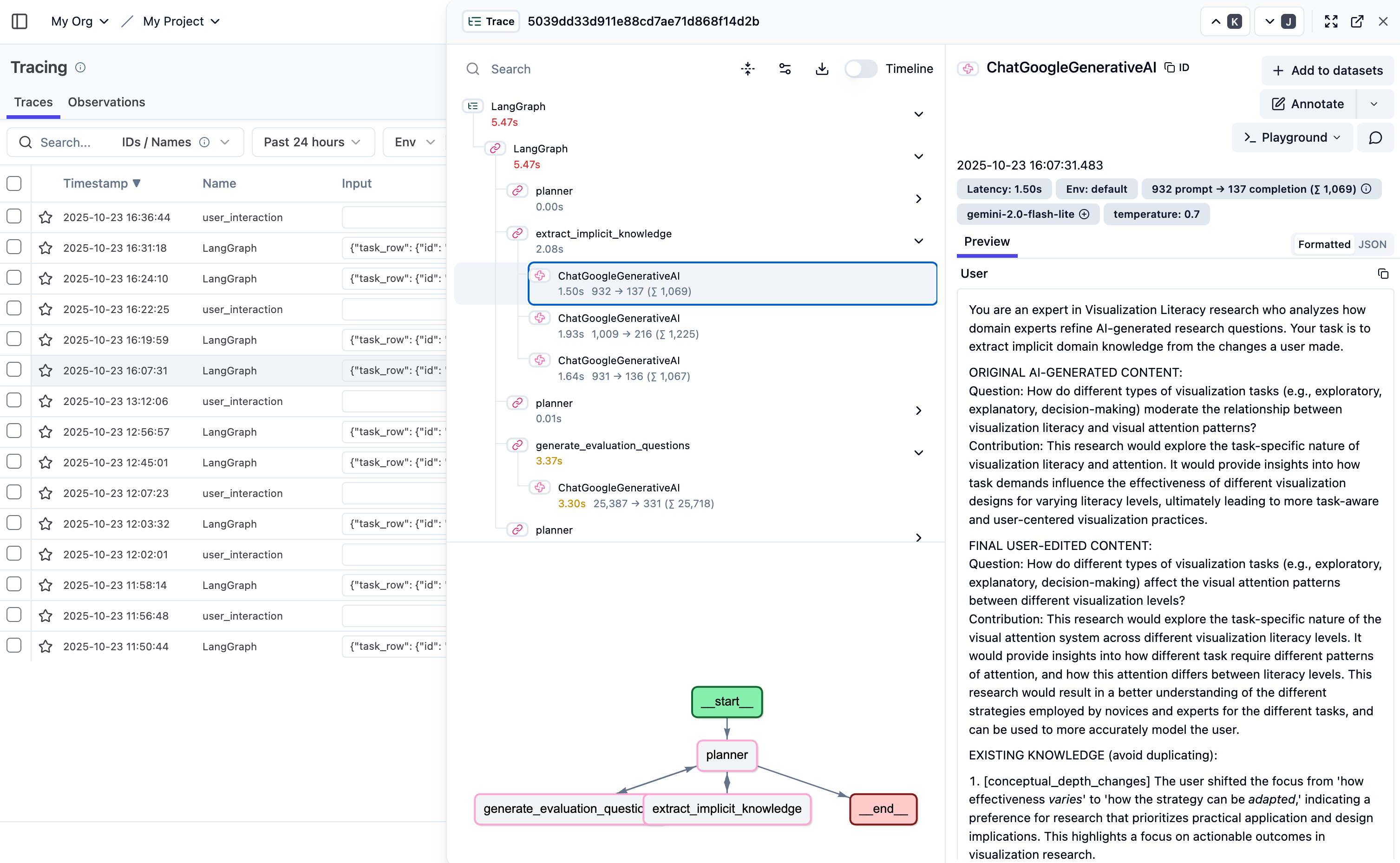}
    \caption{\textbf{Context-mediated domain adaptation workflow tracing.}
    Langfuse tracing demonstrates how the bidirectional learning cycle operates: user modifications flow through the \texttt{extract\_implicit\_knowledge} node (shown processing three user interactions), with extracted knowledge subsequently injected into the \texttt{generate\_evaluation\_questions} node's system prompt.
    Notice how the interface makes the complete knowledge transfer visible, from hierarchical execution flow (center) to detailed prompts and performance metrics (right), enabling validation of our CMDA framework's core claim that user edits systematically enhance AI reasoning.
    }
    \label{fig:seedentia_llm_tracing}
    \Description{A screenshot of the Langfuse tracing interface showing multi-agent workflow execution. The interface has three main sections. The left panel displays a hierarchical trace list with timestamps, showing alternating entries for "user_interaction" and "LangGraph" operations from 2025-10-23. Each LangGraph entry shows JSON input data. The center panel shows a tree structure with performance metrics: "LangGraph" at 5.47s, nested "LangGraph" at 5.47s, "planner" at 0.00s, "extract_implicit_knowledge" at 0.08s, and "ChatGoogleGenerativeAI" at 1.60s with token counts (932 prompt + 137 completion, 1,069 total). Below this is a workflow graph showing connected nodes: "start" at top, flowing to "planner", which branches to "generate_evaluation_questions", "extract_implicit_knowledge", and "end". The right panel shows a detailed view with header information (trace ID, timestamp 2025-10-23 16:07:31.483), model configuration (gemini-2.0-flash-lite, temperature 0.7), and two content sections. The "Preview" tab displays the User prompt describing a visualization literacy research expert task, followed by "ORIGINAL AI-GENERATED CONTENT" and "FINAL USER-EDITED CONTENT" sections showing research questions and contributions about visualization tasks and literacy levels. An "EXISTING KNOWLEDGE" section lists conceptual depth changes about strategic focus shifts in visualization research.}
\end{figure*}

\begin{itemize}
    \item \textbf{Multi-Agent Workflow Tracing}: Figure~\ref{fig:seedentia_llm_tracing} demonstrates the comprehensive tracing capabilities integrated throughout the LangGraph workflow execution.
    The Langfuse interface provides hierarchical visibility into each workflow node, showing the complete execution flow from user interaction through knowledge extraction to subsequent generation enhancement.
    As illustrated in the figure, the \texttt{extract\_implicit\_knowledge} node processes multiple user interactions simultaneously, with each LLM call handling individual modifications to extract domain insights with precise context awareness.

    \item \textbf{Knowledge Lifecycle Monitoring}: The system tracks the complete knowledge lifecycle from extraction through application, providing visibility into knowledge reuse patterns, adaptation effectiveness over time, and the evolution of domain understanding across multiple user sessions.
    The \texttt{implicit\_domain\_knowledge} table enables visualization of clustered extracted knowledge with full provenance tracking through \texttt{source\_question\_ids} arrays, allowing researchers to trace each knowledge entry back to specific user modifications that generated it.
    The selected LLM call in Figure~\ref{fig:seedentia_llm_tracing} displays the complete system prompt, demonstrating how accumulated domain knowledge is injected into generation contexts and how this process is made transparent through comprehensive tracing.

    \item \textbf{Performance and Cost Analysis}: Integrated monitoring captures evaluation metrics in real-time, correlating user interaction patterns with system performance indicators including processing time, model selection, token usage, and estimated costs.
    This enables dynamic analysis of how different participant expertise levels, session lengths, and domain contexts affect adaptation outcomes, providing rich data for both operational optimization and research validation.

    \item \textbf{Provenance and Reproducibility}: All \ac{llm} interactions route through unified functions that preserve generation metadata and semantic relationships, ensuring consistency with our bidirectional semantic links architecture.
    The \texttt{ai\_entity\_metadata} table maintains comprehensive generation context including original prompts, model parameters, and generation code, enabling full reproducibility of generation processes and detailed analysis of how context modifications influence output quality.

\end{itemize}

\subsection{Implementation Status}

Our current implementation provides comprehensive infrastructure for context-mediated domain adaptation, fully realizing the theoretical framework through a working prototype.
The system successfully demonstrates bidirectional semantic links, sophisticated workflow orchestration, comprehensive edit tracking, and automated knowledge extraction with cross-participant learning.
Critical to this success is the integrated monitoring and observability infrastructure that provides transparency into the adaptation process, enabling both operational reliability and systematic research validation of the theoretical framework.

Current limitations include the focus on research question generation tasks, though the architecture supports extension to other content types.
The knowledge extraction relies on LLM-based analysis, which may introduce variability in categorization consistency, though comprehensive monitoring enables detection and analysis of such variations.
Additionally, the system requires domain experts as participants rather than general users, limiting broader applicability.
While the monitoring infrastructure provides comprehensive visibility for researchers and system operators, the accumulated domain knowledge remains largely invisible to end users, limiting their ability to inspect or directly refine the system's learned understanding.
The extensive observability data collected presents opportunities for developing user-facing transparency features in future iterations.

Future development will address evaluation at scale, validation across multiple domains beyond visualization literacy, integration of advanced context management architectures~\cite{du2024survey,zhuang2025selftaught} for handling larger knowledge bases, and collaborative knowledge visualization approaches~\cite{dork2020codesigning,weck2021knowledge} to make the accumulated domain expertise transparent and editable by users.

\subsection{Adapting to New Domains}

The CMDA framework generalizes to other domains through modifications to three components while preserving core multi-agent infrastructure:
The \ac{bdar} pattern, which stores both AI-generated initial states and user-modified final states, is universally applicable.
E.g. for visualization generation, replace \texttt{initial\_question} with \texttt{initial\_visualization\_spec}; for code generation, use \texttt{initial\_code} and \texttt{requirements\_specification}.
The field names change, but the before-and-after comparison mechanism remains identical.
Generation and extraction prompts require domain-specific customization while maintaining structural patterns.
Replace role specifications (e.g., "visualization researcher" → "medical researcher") and adjust knowledge category examples to domain conventions.
The three knowledge categories (domain terminology evolution, methodological refinements, conceptual depth changes) apply across domains but require examples adjusted to domain conventions.
The three interaction modes introduced in section \ref{sec:bidirectional-links} remain applicable, but the user interfaces require domain-specific adaptation.

The Seedentia prototype is available to researchers upon request.
We provide complete source code (Next.js frontend, Python/LangGraph backend), database schemas, Docker configuration, and documentation.
The modular architecture enables straightforward customization of domain-specific components while preserving core CMDA infrastructure.
\section{Evaluation}
\label{sec:evaluation}

We conducted an exploratory proof-of-concept study to assess the feasibility and potential of Context-Mediated Domain Adaptation through research question generation tasks with domain experts in visualization literacy.
This preliminary investigation establishes whether the CMDA mechanisms can capture and apply implicit domain knowledge, providing initial evidence to motivate future controlled validation studies.

\subsection{Method}

\subsubsection{Participants}

We recruited five visualization literacy experts (PhD students and postdoctoral researchers, 1–9 years of experience) with active research backgrounds in visualization, immersive analytics, and applied visualization.
The participant pool represents a range of experience levels from early-career researchers to established postdoctoral researchers, ensuring diverse perspectives on research question quality and domain relevance.

\subsubsection{Materials and Study Design}

Our evaluation employed a sequential knowledge accumulation design where participants processed three academic papers from the visualization literacy domain.
We selected three recent papers focused on visualization literacy: ``Tell Me Without Telling Me: Two-Way Prediction of Visualization Literacy and Visual Attention''~\cite{chang2025telltellingmetwoway}, ``DRIVE-T: A Methodology for Discriminative and Representative Data Viz Item Selection for Literacy Construct and Assessment''~\cite{locoro2025drivetmethodologydiscriminativerepresentative}, and ``Charts-of-Thought: Enhancing LLM Visualization Literacy Through Structured Data Extraction''~\cite{das2025chartsofthoughtenhancingllmvisualization}.
The complete evaluation data, including participant responses, extracted domain knowledge entries, and analysis scripts, are available at our anonymized OSF repository.\footnote{\url{https://osf.io/84f3s/overview?view_only=6b4f191fa0d341c4803bf53d3229e3fd}}

The critical aspect of our study was sequential learning:
participants used the system one after another, with each benefiting from knowledge accumulated within the system through the interactions of previous participants.
Specifically, Participant 1 receives baseline AI output generated without domain knowledge, Participant 2 benefits from knowledge extracted from Participant 1's modifications across all three papers, Participant 3 leverages accumulated knowledge from both P1 and P2's complete sessions, Participant 4 receives the full accumulated knowledge from all three prior participants, and Participant 5 benefits from the complete accumulated knowledge from all four previous participants.
Importantly, the system learns not just between participants but also within each participant's session—knowledge extracted from editing the first paper improves generation for the second and third papers within the same session.
This design enables us to observe how accumulated domain expertise affects initial generation quality and user editing effort both within and across participant sequences.

\subsubsection{Procedure}

For each paper, participants reviewed the abstract and full text, then received three AI-generated research questions with contribution summaries.
They rated each question on a 1–5 Likert scale before editing and refined content using direct manipulation or prompt-based regeneration until satisfied.
All edits were logged in real time, and extracted knowledge was used to improve subsequent generations.
Each session lasted approximately 45-60 minutes, with participants encouraged to think aloud during the refinement process.

\subsubsection{Metrics and Analysis}

We track five primary metrics to assess context-mediated adaptation.
\textbf{Edit distance} measures character-level changes between initial and final versions, with decreasing distances across participants indicating improved initial generation quality.
\textbf{Initial generation quality} uses participant ratings (1-5 scale) before any edits, with increasing ratings demonstrating successful knowledge transfer.
\textbf{Time-to-completion} captures session duration from generation to participant satisfaction.
\textbf{Knowledge accumulation} counts unique domain knowledge entries extracted per participant, revealing contribution patterns.
\textbf{Interaction mode usage} tracks the frequency of direct edits versus prompt-based regeneration, indicating participant preferences and system behavior.

The system automatically extracts domain knowledge from user modifications, categorizing insights into methodological refinements, conceptual depth changes, and domain terminology evolution.
Knowledge saturation emerges when extraction rates decrease and existing knowledge reuse increases, indicating comprehensive domain coverage.

\subsection{Analysis Framework and Threats to Validity}

Our analysis was guided by four hypotheses based on the theoretical framework.
We examined whether edit distances decreased across participant positions as the system learned domain conventions.
We measured if time-to-completion decreased for later participants who benefited from accumulated knowledge.
We tracked whether initial generation quality ratings increased across the participant sequence, demonstrating successful knowledge transfer.
Finally, we investigated whether knowledge saturation emerged, with fewer novel knowledge items extracted from later participants.

Several threats to validity must be acknowledged.
Internal validity concerns include participant fatigue across three papers and potential learning effects within sessions.
External validity is limited by our focus on visualization literacy, with generalization to other domains remaining untested.
Construct validity depends on subjective quality assessments that may vary across participants.
We address these threats through standardized evaluation criteria, randomized paper presentation order where feasible, and detailed interaction logging for post-hoc analysis of confounding factors.

\subsection{Results}

Our evaluation yielded rich data on the feasibility of context-mediated domain adaptation through 47 refined research questions and 46 extracted domain knowledge entries across five participants who used the system sequentially, with each participant benefiting from the accumulated knowledge of all previous users.

\subsubsection{Quantitative Results}

We begin by examining how participants interacted with the system to understand engagement patterns and initial quality perception.
Table~\ref{tab:participation_data} reveals distinct interaction strategies across participants in our sequential design.

\begin{table*}[htb]
  \centering
  \caption{\textbf{Participation data showing temporal patterns in domain adaptation.}
    Key insight: Quality ratings improve across sequential participants (P1: 2.67-3.0 → P5: 4.0-4.33), demonstrating progressive knowledge accumulation.
    Duration (minutes), quality ratings (1=poor to 5=excellent), and edit counts reveal how different participants engaged with the system.
  }
  \label{tab:participation_data}
  \small
  \begin{tabular}{@{}p{0.8cm}p{3.8cm}c c c c c@{}}
  \toprule
  \textbf{Part.} & \textbf{Paper} & \makecell{\textbf{Duration} \\ \small{minutes}} & \makecell{\textbf{\contextedittext{Quality Rating}} \\ \small{1=poor, 5=excellent}} & \makecell{\textbf{\directedittext{Direct Edits}} \\ \small{total sum}} & \makecell{\textbf{\promptedittext{Prompt Edits}} \\ \small{total sum}} \\
  \midrule
  P1 & Tell Me Without Telling Me & 9.0 & {\contextedittext{2.67}} & {\directedittext{4}} & {\promptedittext{0}} \\
     & DRIVE-T Methodology        & 5.1 & {\contextedittext{3.0}}  & {\directedittext{1}} & {\promptedittext{0}} \\
     & Charts-of-Thought          & 4.8 & {\contextedittext{3.0}}  & {\directedittext{1}} & {\promptedittext{0}} \\
  \midrule
  P2 & Tell Me Without Telling Me & 32.5 & {\contextedittext{2.33}} & {\directedittext{5}} & {\promptedittext{1}} \\
     & DRIVE-T Methodology        & 18.8 & {\contextedittext{3.67}} & {\directedittext{3}} & {\promptedittext{1}} \\
     & Charts-of-Thought          & 14.2 & {\contextedittext{2.67}} & {\directedittext{3}} & {\promptedittext{1}} \\
  \midrule
  P3 & Tell Me Without Telling Me & 16.6 & {\contextedittext{3.0}}  & {\directedittext{4}} & {\promptedittext{0}} \\
     & DRIVE-T Methodology        & 10.2 & {\contextedittext{2.67}} & {\directedittext{2}} & {\promptedittext{0}} \\
     & Charts-of-Thought (session 1) & 8.0  & {\contextedittext{3.67}} & {\directedittext{1}} & {\promptedittext{3}} \\
     & Charts-of-Thought (session 2) & 4.1  & {\contextedittext{3.33}} & {\directedittext{0}} & {\promptedittext{2}} \\
  \midrule
  P4 & Tell Me Without Telling Me & 6.9 & {\contextedittext{4.33}} & {\directedittext{0}} & {\promptedittext{0}} \\
     & DRIVE-T Methodology        & 4.8 & {\contextedittext{3.67}} & {\directedittext{1}} & {\promptedittext{0}} \\
     & Charts-of-Thought          & 5.0 & {\contextedittext{4.33}} & {\directedittext{0}} & {\promptedittext{1}} \\
  \midrule
  P5 & Tell Me Without Telling Me & 4.0 & {\contextedittext{3.67}} & {\directedittext{1}} & {\promptedittext{0}} \\
     & DRIVE-T Methodology        & 4.0 & {\contextedittext{3.0}}  & {\directedittext{1}} & {\promptedittext{0}} \\
     & Charts-of-Thought          & 2.6 & {\contextedittext{4.0}}  & {\directedittext{0}} & {\promptedittext{0}} \\
  \bottomrule
  \end{tabular}
\end{table*}

The temporal patterns suggest potential efficiency gains from knowledge accumulation, although time was not a primary metric and individual differences likely contribute. 
P1 averaged 6.3 minutes per paper (18.9 total), whereas P5 completed all three papers in 10.6 minutes. 
Quality ratings show a similar pattern: later participants achieved higher baseline ratings with minimal intervention (e.g., P4: 4.11 average; P5: 3.56 with only two direct edits), compared to P1's lower baseline (2.89) despite more edits. 
Without a control condition, these trends cannot be causally attributed to CMDA rather than participant characteristics or paper difficulty.

Edit behavior did not decrease monotonically. 
Despite benefiting from 22 prior knowledge entries, P3 showed the highest editing activity (7 direct, 5 prompt), suggesting that improved baseline generations may enable deeper, more sophisticated refinement rather than simply reducing effort. 
Interaction logs reflect complementary roles of the two modalities: direct edits were used primarily for minor adjustments (28 instances), while prompt-based regeneration supported fundamental restructuring (7 instances). 
P3's double evaluation of \textit{Charts-of-Thought} further illustrates within-user adaptation dynamics: the second session was faster (4.1 vs.\ 8.0 minutes) with comparable quality (3.33 vs.\ 3.67), consistent with both system learning from earlier edits and participant familiarity. 
P2's longer duration (65.5 minutes) included interruptions, making time less comparable across participants; overall, participants were not incentivized to optimize for speed, so timing should be interpreted as a secondary indicator.

Across the sequence, baseline quality ratings increased from early participants (P1: 2.67–3.0) to later participants (P4–P5: 3.0–4.33), consistent with progressive knowledge accumulation. 
While the small sample (n=5; 47 rated questions) and lack of a control condition preclude causal claims, the pattern provides preliminary evidence that CMDA may contribute to improved initial generation quality and motivates future controlled validation.

Apart from the engagement with the system, we now analyze what specific modifications were made.
Table~\ref{tab:participation_edits} quantifies these modifications through character-level edit distances and field counts.

\begin{table*}[htb]
  \centering
  \caption{\textbf{Character-level edit distances and field modifications by participant.}
  Edit distances quantify character-level changes between initial AI generation and final versions across research questions (Q) and contribution summaries (C).
  Notice how P3's extensive modifications (2893 total characters) coupled with high quality ratings suggest substantive refinements rather than surface corrections, while P4's minimal edits reflect satisfaction with generation quality.
  }
  \label{tab:participation_edits}
  \small
  \begin{tabular}{@{}p{0.8cm}p{3.8cm}c c @{}}
  \toprule
  \textbf{Part.} & \textbf{Paper} & \makecell{\textbf{Edit Distance} \\ \small{Q=Question, C=Contribution}} & \makecell{\directedittext{\textbf{Edited fields}} \\ amount per paper} \\
  \midrule
  P1 & Tell Me Without Telling Me &  {Q: 71   -  C: 22 } & \directedittext{6} \\
     & DRIVE-T Methodology        &  {Q: 19   -  C: 0  } & \directedittext{4} \\
     & Charts-of-Thought          &  {Q: 50   -  C: 0  } & \directedittext{4} \\
  \midrule
  P2 & Tell Me Without Telling Me &  {Q: 103  -  C: 121} & \directedittext{3}  \\
     & DRIVE-T Methodology        &  {Q: 82   -  C: 63 } & \directedittext{1} \\
     & Charts-of-Thought          &  {Q: 83   -  C: 74 } & \directedittext{1} \\
  \midrule
  P3 & Tell Me Without Telling Me &  {Q: 43   -  C: 41 } & \directedittext{4} \\
     & DRIVE-T Methodology        &  {Q: 49   -  C: 17 } & \directedittext{2} \\
     & Charts-of-Thought          &  {Q: 197  -  C: 201} & \directedittext{6} \\
     & Charts-of-Thought          &  {Q: 143  -  C: 140} & \directedittext{4} \\
  \midrule
  P4 & Tell Me Without Telling Me &  {Q: 0    -  C: 0  } & \directedittext{0}  \\
     & DRIVE-T Methodology        &  {Q: 10   -  C: 0  } & \directedittext{1} \\
     & Charts-of-Thought          &  {Q: 22   -  C: 61 } & \directedittext{1} \\
  \midrule
  P5 & Tell Me Without Telling Me &  {Q: 0    -  C: 0  } & \directedittext{1}  \\
     & DRIVE-T Methodology        &  {Q: 0    -  C: 0  } & \directedittext{1} \\
     & Charts-of-Thought          &  {Q: 0    -  C: 0  } & \directedittext{0} \\
  \bottomrule
  \end{tabular}
\end{table*}

Edit distance did not decrease monotonically as knowledge accumulated, indicating that adaptation does not simply reduce editing effort.
Instead, modification intensity varied widely, with P3 producing the most edits despite benefiting from 22 prior knowledge entries.
This suggests that accumulated knowledge enables deeper expert refinement rather than discouraging engagement.

P3's second evaluation of the \textit{Charts-of-Thought} paper still involved substantial edits, but these reflected new and complementary refinements because earlier insights had already been integrated.
This pattern indicates that CMDA supports iterative quality improvement rather than redundant correction.

Participants differed in editing focus, with some prioritizing research question framing and others refining contribution context and impact.
Over time, research questions became longer and more conceptually rich, incorporating methodological detail, theoretical framing, and applied considerations such as accessibility.

Edits encoded meaningful domain expertise rather than superficial corrections.
From these refinements, we extracted 46 unique domain knowledge entries, showing a strong positive relationship between editing activity and knowledge extraction (slope = 0.78).
This confirms that user edits act as a high-signal channel for implicit domain knowledge transfer, supporting ongoing system adaptation.

Table~\ref{tab:participation_knowledge_data} breaks down how this knowledge is distributed across participants and categories, revealing the depth of expertise captured.

\begin{table*}[htb]
  \centering
  \caption{\textbf{Distribution of extracted knowledge entries by participant and category.}
    Notice how P3 contributed the most knowledge entries (n=20) despite benefiting from accumulated knowledge, demonstrating that the system enables rather than replaces expert contribution.
    The uneven distribution (P1: 6, P2: 16, P3: 20, P4: 3, P5: 1) reflects varying engagement levels and expertise expression styles.
  }
  \label{tab:participation_knowledge_data}
  \small
  \begin{tabular}{@{}p{0.8cm}| c | c c c | c@{}}
  \toprule
  \textbf{Part.} & \makecell{\directedittext{\textbf{Edited fields}} \\ total amount} & \makecell{\textbf{Conceptual} \\ \textbf{Depth Changes}} & \makecell{\textbf{Domain Terminology} \\ \textbf{Evolution}} & \makecell{\textbf{Methodological} \\ \textbf{Refinements}} & {\contextedittext{\textbf{Total}}} \\
  \midrule
  P1 & \directedittext{14} & 2 & 3 & 1 & {\contextedittext{6}} \\
  \midrule
  P2 & \directedittext{5}  & 9 & 2 & 5 & {\contextedittext{16}} \\
  \midrule
  P3 & \directedittext{15} & 12 & 5 & 3 & {\contextedittext{20}} \\
  \midrule
  P4 & \directedittext{3}  & 2 & 0 & 1 & {\contextedittext{3}} \\
  \midrule
  P5 & \directedittext{2}  & 1 & 0 & 0 & {\contextedittext{1}} \\
  \bottomrule
   & \directedittext{\textbf{39}}  & \textbf{26} & \textbf{10} & \textbf{10} & {\contextedittext{\textbf{46}}}
  \end{tabular}
\end{table*}

The distribution challenges conventional assumptions about knowledge saturation.
P3 contributed the most knowledge (20 entries) despite benefiting from 22 previously extracted entries, particularly emphasizing conceptual depth changes (12 entries).
The double evaluation of Charts-of-Thought paper by P3 provides evidence for the system's learning capability: the second session's knowledge entries were complementary rather than duplicative, as the system had already integrated learnings from the first session into its generation baseline.
This pattern suggests bidirectional learning—the system improves from edits while simultaneously enabling experts to explore different dimensions of quality in subsequent interactions.
This counterintuitive pattern suggests that prior domain knowledge creates a foundation for increasingly sophisticated expert contributions rather than reaching a plateau.

As shown in Table~\ref{tab:participation_knowledge_data}, the system categorized 46 unique knowledge entries into three primary types introduced in Section~\ref{sec:knowledge-extraction}: Conceptual Depth Changes (56.5\%, n=26) representing expansions in scope and theoretical frameworks, Domain Terminology Evolution (21.7\%, n=10) reflecting refinements in technical language, and Methodological Refinements (21.7\%, n=10) capturing improvements in assessment techniques.

The predominance of conceptual depth changes reveals that experts primarily expand the system's understanding of research implications rather than correcting surface-level language.
This distribution aligns with our framework's goal of capturing tacit expert knowledge that extends beyond vocabulary corrections, showing that the system captures multiple dimensions of domain expertise with the majority of contributions advancing conceptual understanding rather than making cosmetic corrections.

\subsubsection{Qualitative Findings}

Having quantified the patterns of interaction, modification, and knowledge extraction, we now examine specific examples to illustrate how domain expertise manifests through user modifications.
We present these examples in order of increasing depth, from surface-level terminology refinements to fundamental conceptual expansions.

\paragraph{Surface-Level Refinements: Precision in Language}

The most basic level of extracted knowledge involves terminology refinements that enhance precision without altering fundamental concepts.
For example, based on user interaction the system generated this insight:
\begin{hquote}{contextgeneration}
  \textit{``The user removed the word `granularity' from the question, suggesting a preference for more concise and direct language when framing research questions about assessment methods.''}
\end{hquote}
While such edits might seem cosmetic, they teach the system domain-specific communication norms—what constitutes clear, professional discourse in visualization research.

More substantively, participants consistently refined vague terms to precise constructs:
\begin{hquote}{contextgeneration}
  \textit{``The user replaced broader task categories (exploratory, explanatory, decision-making) with more specific ones (lookup, search, filtering), suggesting a preference for tasks more directly tied to interaction and information retrieval within visualizations, which is a common focus in usability and attention research.''}
\end{hquote}
This transformation from generic categorizations to specific interaction patterns represents essential domain vocabulary that distinguishes expert discourse from general descriptions.

At a deeper level, participants introduced methodological refinements that expand how the system conceptualizes research approaches.
A particularly impactful example involves multimodal assessment:
\begin{hquote}{contextgeneration}
  \textit{``The refined question emphasizes the use of 'physiological signals' alongside visual attention and cognitive factors, suggesting a shift towards a more holistic and potentially multimodal approach to assessing visualization literacy, perhaps incorporating biofeedback or real-time user state monitoring.''}
\end{hquote}
This modification doesn't merely correct terminology but introduces an entirely new methodological paradigm—teaching the system that modern visualization assessment extends beyond traditional cognitive metrics.

The most profound knowledge contributions involve conceptual expansions that fundamentally reshape how the system approaches research questions.
One notable example introduced accessibility considerations:
\begin{hquote}{contextgeneration}
  \textit{``The user expanded the potential impact of the research to include assisting 'low-vision users with visual analytics tasks', which highlights the importance of accessibility considerations when designing and evaluating interactive visualizations and LLM-based solutions.''}
\end{hquote}
This modification transcends correction—it teaches the system that visualization research must consider diverse user populations and inclusive design principles.

Participant feedback strongly validates the need for context-mediated domain adaptation.
A critical theme emerged regarding the essential role of domain expertise in refining AI-generated content, with one participant emphasizing that 
\begin{hquote}{directedit}
  \textit{``
  novelty stems from creating your own ideas. So after the first creation by LLMs, it is always important to think about it yourself as well.
  ''}
\end{hquote}
This observation underscores that while AI systems can provide initial scaffolding, meaningful research contributions require domain experts to inject their specialized knowledge and critical thinking.

Participants consistently reported that AI-generated research questions were ``superficial or not that interesting on their own'' but ``could be a good starting point,'' highlighting the gap between generic AI output and domain-specific requirements.
Furthermore, participants expressed desire for systems that could learn from their expertise, with one noting the tool would be more useful
\begin{hquote}{directedit}
  \textit{``
  if it would also give feedback on my direct edits, suggesting changes and formulations that capture things I did not think of.
  ''}
\end{hquote}

The consistent preference for direct editing over prompting—with participants noting that 
\begin{hquote}{directedit}
  \textit{``
  improving the question myself was faster
  ''}
\end{hquote}
—demonstrates that current AI systems lack the domain-specific knowledge required for specialized tasks.

To assess the quality and actionability of our knowledge extraction, we examined five representative entries that illustrate both the strengths and limitations of automated domain knowledge capture.
The most impactful entry (\textbf{conceptual\_depth\_changes}) expanded research scope to include 
\begin{hquote}{contextgeneration}
  \textit{``
  The user expanded the potential impact of the research to include assisting `low-vision users with visual analytics tasks', which highlights the importance of accessibility considerations when designing and evaluating interactive visualizations and LLM-based solutions.
  ''}
\end{hquote}
This illustrates how expert modifications introduce critical accessibility considerations absent from initial AI generation—knowledge that reshapes how the system approaches visualization research questions.
Similarly valuable is the shift (\textbf{domain\_terminology\_evolution}) from broad task categories to specific interaction patterns:
\begin{hquote}{contextgeneration}
  \textit{``
  The user replaced broader task categories (exploratory, explanatory, decision-making) with more specific ones (lookup, search, filtering), suggesting a preference for tasks more directly tied to interaction and information retrieval within visualizations, which is a common focus in usability and attention research.
  ''}
\end{hquote}
This teaches the system precise terminology that distinguishes expert discourse from generic descriptions.
The emphasis on multimodal assessment (\textbf{methodological\_refinements}) reveals how experts push beyond conventional metrics:
\begin{hquote}{contextgeneration}
  \textit{``
  The refined question emphasizes the use of `physiological signals' alongside visual attention and cognitive factors, suggesting a shift towards a more holistic and potentially multimodal approach to assessing visualization literacy, perhaps incorporating biofeedback or real-time user state monitoring.
  ''}
\end{hquote}

\subsubsection{Synthesis of Findings}

Our evaluation reveals a coherent narrative of how context-mediated domain adaptation transforms user corrections into persistent system improvements.
The progression from HOW participants interact, through WHAT they modify, to WHY they make changes illustrates the feasibility of bidirectional semantic links in capturing and propagating domain expertise.

A notable pattern is that accumulated knowledge creates a virtuous cycle rather than reaching saturation.
P3's extensive contributions (20 knowledge entries) despite benefiting from 22 prior entries illustrates this pattern.
The double evaluation of Charts-of-Thought paper particularly illustrates the system's learning dynamics: in the first session, P3 refined methodological aspects and terminology precision, while the second session's improved baseline (from incorporating first session edits) enabled P3 to focus on different dimensions such as theoretical frameworks and cross-domain implications.
This suggests that context-mediated adaptation enables iterative quality improvement rather than simple error correction.
Later participants achieve higher quality ratings with less editing effort—P4's 4.11 average rating with minimal intervention contrasts with P1's 2.89 rating despite extensive editing (6 direct edits).
While time was not a primary metric, P3's double evaluation provides interesting context: the second session took 4.1 minutes versus 8.0 minutes initially, though this likely reflects both system learning and participant familiarity with the paper.
Modifications become more sophisticated over time—question length increases 35\% as the system learns to generate more nuanced formulations, while edit patterns shift from surface corrections to deep conceptual refinements.
The knowledge extracted evolves from basic terminology (early participants) to complex conceptual frameworks (later participants), with 56.5\% of entries representing conceptual depth changes that fundamentally reshape the system's approach.
Returning to our initial expectations, the results provide preliminary observations:

\begin{itemize}
\item[H1]\textit{Decreasing edit distances:} Pattern not observed uniformly—P3's extensive modifications suggest that knowledge accumulation may enable deeper engagement rather than simply reducing effort.

\item[H2]\textit{Decreasing time-to-completion:} Pattern consistent with expectations—average session time decreased from P1's 6.3 minutes per paper to P4's 5.6 minutes, though individual differences and paper familiarity effects cannot be ruled out.

\item[H3]\textit{Increasing quality ratings:} Pattern observed—42\% improvement from P1 to P4 is consistent with knowledge transfer, though baseline condition needed to establish causality.

\item[H4]\textit{Knowledge saturation:} Pattern not observed—the system showed no signs of saturation within this limited study, with P3 contributing substantial knowledge despite prior accumulation.

\end{itemize}
 
\section{Discussion}

Our evaluation provides preliminary evidence that context-mediated domain adaptation can bridge the gap between generic AI capabilities and domain-specific expertise requirements.
The 46 extracted knowledge entries across five sequential participants demonstrate the feasibility of treating user modifications as implicit specifications for system behavior adaptation.

\subsection{Engineering Contributions}

Our work provides concrete architectural patterns for building adaptive human-AI collaborative systems.
The framework implements bidirectional semantic links between user interactions and system reasoning through persistent knowledge graph structures that maintain relationships between generated artifacts and extracted domain patterns.
Multi-layered context assembly pipelines aggregate knowledge at user, project, and global scopes, while event-driven extraction mechanisms automatically categorize modifications into the three knowledge categories: domain terminology evolution, methodological refinements, and conceptual depth changes.
This architecture provides reusable patterns for any domain requiring continuous adaptation to user expertise.

The system's three interaction modes establish engineering best practices for accommodating varying levels of user intervention while maintaining semantic connections throughout.
Our instrumentation framework captures edit distances at character and word levels, tracks knowledge saturation through generation-versus-reuse ratios, and enables longitudinal analysis of accumulation patterns.
These metrics provide quantitative methods for evaluating knowledge transfer effectiveness in adaptive systems.

The database architecture implements normalized knowledge representation that separates domain insights from specific instances, with versioning and provenance tracking maintaining complete evolution history.
Scoped access patterns enable privacy-preserving knowledge sharing, while LLM-agnostic storage ensures portability across different AI backends.
The complete knowledge lifecycle—from extraction through categorization to integration—follows systematic processes that analyze user modifications, classify insights into actionable types, and incorporate extracted knowledge into system prompts.

These contributions directly address key engineering challenges in interactive AI systems.
Automatic extraction and categorization eliminate manual prompt engineering, enabling scalability across domains without extensive configuration.
Component separation ensures maintainability through independent subsystem evolution.
Continuous knowledge accumulation allows systems to evolve with their user communities, reducing manual customization burden.
The domain-agnostic patterns provide reusable components for diverse application contexts.

\subsection{Limitations}

Our approach faces several important constraints that bound its applicability.
The current implementation focuses on research question generation within visualization literacy, and while the architecture supports extension, the extraction patterns remain specialized for academic sensemaking tasks.
With only five participants in our visualization literacy evaluation and 46 knowledge entries, we provide proof-of-concept validation but cannot demonstrate comprehensive domain coverage or scalability to larger communities.
Transfer to significantly different domains would require adapting the categorization and extraction mechanisms.

The system depends on domain experts capable of providing meaningful corrections.
General users without sufficient expertise may not produce corrections that yield useful learning signals, limiting applicability beyond expert communities.
Our approach works best for structured, text-based artifacts with clear editability boundaries; complex multimedia content or real-time collaborative tasks may strain the semantic linking mechanisms.
We observed knowledge transfer across participants but not long-term retention or drift, and subjective quality ratings may not fully capture the nuanced improvements in generated content.

Our knowledge extraction relies on carefully crafted prompts that may not be optimal across all scenarios.
The quality and consistency of extracted knowledge varies significantly based on the LLM model used—we primarily tested with Gemini 2.0 Flash Lite—and different models may produce substantially different categorizations.
This model-dependence introduces variability that could affect reproducibility and cross-system compatibility.

Despite these limitations, the consistent patterns across multiple metrics—quality ratings, edit distances, knowledge extraction, question complexity, and temporal efficiency—provide converging evidence that context-mediated domain adaptation can bridge the gap between generic AI capabilities and domain-specific requirements.
Participant feedback validates our approach: experts explicitly desired systems that learn from their corrections to improve future generations, which context-mediated domain adaptation delivers through systematic capture of edit patterns and domain conventions.

\subsection{Future Work}

Our evaluation reveals that accumulated knowledge enables rather than replaces expert contribution, with the 0.78 correlation between editing activity and knowledge extraction demonstrating effective implicit knowledge transfer.
This finding suggests new research directions for sustainable knowledge management in adaptive AI systems.

Knowledge lifecycle management presents immediate challenges—as domains evolve, accumulated patterns may become outdated.
Future systems need aging mechanisms that preserve foundational insights while deprioritizing obsolete patterns, validation methods to distinguish expert corrections from individual preferences, and consolidation strategies for conflicting knowledge entries.

The current three-category taxonomy (conceptual depth, terminology, methodology) may not optimally capture all domain expertise forms.
Research should explore automatic category discovery through unsupervised clustering and self-organizing taxonomies that adapt to different domains.
Enterprise deployment requires governance mechanisms for knowledge sharing, privacy-preserving federated learning, and role-based access controls.

Cross-domain transfer remains unexplored—can knowledge from research questions generalize to related tasks?
Knowledge graph visualization approaches~\cite{li2023knowledge,li2021kg4vis} could enable users to inspect and refine accumulated domain understanding, providing transparency into how their corrections shape system behavior. 

\begin{acks}
    This work was supported partly by Villum Investigator grant VL-54492 by Villum Fonden.
    Any opinions, findings, and conclusions expressed in this material are those of the authors and do not necessarily reflect the views of the funding agency.
\end{acks}

\bibliographystyle{ACM-Reference-Format}
\bibliography{seedentia}

\end{document}